\documentclass[twocolumn,aps,prb,superscriptaddress,groupedaddress,showpacs]{revtex4}
\usepackage{epsfig}
\begin{document}
\title{
\Large\bf Field theory of bicritical and tetracritical points. III. Relaxational dynamics including
conservation of magnetization (Model C)}
\author{R. Folk}\email{folk@tphys.uni-linz.ac.at}
\affiliation{Institute for Theoretical Physics, Johannes Kepler
University Linz, Altenbergerstrasse 69, A-4040, Linz, Austria}
 \author{Yu. Holovatch}\email[]{hol@icmp.lviv.ua}
 \affiliation{Institute for Condensed Matter Physics, National
Academy of Sciences of Ukraine, 1~Svientsitskii Str., UA--79011
Lviv, Ukraine}\affiliation{Institute for Theoretical Physics, Johannes Kepler
University Linz, Altenbergerstrasse 69, A-4040, Linz, Austria}
  \author{G. Moser}\email[]{guenter.moser@sbg.ac.at}
\affiliation{Department for Material Research and Physics, Paris Lodron University
Salzburg, Hellbrunnerstrasse 34, A-5020 Salzburg, Austria}
\date{\today}
\begin{abstract}
We calculate the relaxational dynamical critical behavior of systems of
$O(n_\|)\oplus O(n_\perp)$ symmetry including conservation of magnetization by renormalization group (RG) theory
within the minimal subtraction scheme in two loop order.
Within the stability region of the Heisenberg fixed point and the biconical fixed point strong dynamical scaling holds
with the  asymptotic dynamical critical exponent  $z=2\phi/\nu-1$ where $\phi$ is the crossover exponent and
$\nu$ the exponent of the correlation length. The critical dynamics at $n_\|=1$ and $n_\perp=2$ is governed by a small
dynamical transient exponent leading to  nonuniversal nonasymptotic dynamical behavior.
This  may be seen e.g. in the temperature dependence of the magnetic transport coefficients.
\end{abstract}
\maketitle
\section{Introduction}
In two preceding papers\cite{partI,partII} we have considered the critical statics and relaxational dynamics
of $O(n_\|)\oplus O(n_\perp)$  physical systems near the multicritical point where the two
phase transition lines of the system corresponding to $O(n_\|)$-symmetry and $O(n_\perp)$-symmetry  meet.
The space of the order parameter (OP) dimensions $n_\|$ and $n_\perp$  decomposes in regions where the multicritical behavior
is described by different fixed points (FPs) -  the $O(n=n_\|+n_\perp)$- isotropic FP, the biconical FP,
the decoupling FP - and  a region where no stable FP is found (run away region) (see Fig 1 in Ref.\cite{partI}).
In the resummed two loop order field theoretic treatment it was found that for integer values of $n_\|$ and $n_\perp$
the biconical FP is stable only for a system with $n_\|=1$, $n_\perp=2$, and its symmetric counterpart. For specific initial
conditions  of the nonuniversal parameters of the system  also the $O(n=n_\|+n_\perp)$- isotropic FP (Heisenberg FP) might be reached.
Such a system is physically represented by an antiferromagnet in an external  magnetic field. The two cases mentioned 
above correspond to tetracritical and bicritical multicritical points correspondingly. If no FP is reached the multicritical
point might be of first order, i.e. a triple point.

The dynamics of the antiferromagnet in a magnetic field is quite
complicated and the equations of motion have been formulated for the
slow densities by Dohm and Janssen\cite{dohmjanssen77}. These Eqs.
contain reversible and irreversible coupling terms between the OPs
(the  components of the staggered magnetization parallel and
perpendicular to the magnetic field) and one conserved density (the
parallel component of the magnetization). In fact there is a second
conserved density  (CD) - the energy density - which in general
should be  taken into account but it will not be included here since
in two loop order the specific heat exponent at the biconical FP
turned out to be negative \cite{rem1} for the case $n_\|=1$ and
$n_\perp=2$. Concerning the new static results\cite{partI} a
simplified dynamical model\cite{dohmjanssen77} has been
reconsidered\cite{partII} consisting of two relaxational equations
for the two OP. The timescale ratio $v$ between the two relaxation
rates $\Gamma_\|$ and  $\Gamma_\perp$ introduces a very small
dynamical transient since the dynamical FP lies very near to the
stability boundary separating strong and weak dynamical scaling.
The strong dynamical scaling FP is governed by  $v^\star$ finite
and different from zero whereas the weak dynamical scaling FP by
$v^\star=0\, \mbox{or}\,\infty$ correspondingly.

A further step to the complete model is to include the diffusive dynamics of the slow CD leading to a
model C like extension. In this extended model a new timescale ratio appears defined by the ratio of one of the
OP-relaxation rate to the kinetic coefficient $\lambda$ of the conserved density $m$. This model has been studied in one loop order in
Refs \cite{dohmjanssen77,dohmKFA,dohmmulti83} taking into  account only a part of dynamical two
loop order terms and one loop statics. Here we present a complete two loop order calculations.

The inclusion of further densities beside the OP makes it necessary to extend the static functional of the usual
$\phi^4$-theory, although the OP alone would be sufficient to describe the static
critical behavior. Such an extended static functional for isotropic systems ($O(n)$
symmetry) with short range interaction has the form \cite{hahoma74,fomoprl03}
\begin{eqnarray}\label{hmodc}
{\cal H}^{(C)}\!=\!\int\! d^dx\Bigg\{\frac{1}{2}\mathring{\tilde{r}}\vec{\phi}_0
\cdot\vec{\phi}_0+\frac{1}{2}\sum_{i=1}^n\nabla_i\vec{\phi}_0\cdot
\nabla_i\vec{\phi}_0+\frac{1}{2}m_0^2  \nonumber \\
+\frac{\mathring{\tilde{u}}}{4!}\Big(\vec{\phi}_0\cdot\vec{\phi}_0\Big)^2
+\frac{1}{2}\mathring{\gamma}m_0\vec{\phi}_0\cdot\vec{\phi}_0
-\mathring{h}m_0 \Bigg\}\ .
\end{eqnarray}
Here the order parameter $\vec{\phi}_0\equiv \vec{\phi}_0(x)$ is
assumed to be a $n$-component real vector, the symbol $\cdot$
denotes the scalar product. The secondary density $m_0\equiv m_0(x)$
is considered as a scalar quantity and $\mathring{h}$ is the
conjugated field to $m_0$. It is chosen to have a vanishing
average value $\langle m_0\rangle=0$. Within statics, the above
functional is equivalent to the
Ginzburg-Landau-Wilson(GLW)-functional
\begin{eqnarray}\label{hglw}
{\cal H}_{GLW}\!=\!\int\! d^dx\Bigg\{\frac{1}{2}\mathring{r}\vec{\phi}_0
\cdot\vec{\phi}_0+\frac{1}{2}\sum_{i=1}^n\nabla_i\vec{\phi}_0\cdot
\nabla_i\vec{\phi}_0  \nonumber \\
+\frac{\mathring{u}}{4!}\Big(\vec{\phi}_0\cdot\vec{\phi}_0\Big)^2 \Bigg\} \ ,
\end{eqnarray}
where $\mathring{r}$ is proportional to
the temperature distance to the critical point and $\mathring{u}$ is
the fourth order coupling in which perturbation expansion is usually
performed. The GLW-functional (\ref{hglw}) is obtained by integrating out the
CD, which appears only in Gaussian order in (\ref{hmodc}), in
the corresponding partition function.
The parameters $\mathring{\tilde{r}}$, $\mathring{\tilde{u}}$ and
$\mathring{\gamma}$ in (\ref{hmodc}) and,
$\mathring{r}$ and $\mathring{u}$ in (\ref{hglw}) are
related by
\begin{equation}\label{rurel}
\mathring{r}=\mathring{\tilde{r}}+\mathring{\gamma}\mathring{h} \ , \qquad
\mathring{u}=\mathring{\tilde{u}}-3\mathring{\gamma}^2 \ .
\end{equation}
The extended static functional appears in the driving force of the equations of motion for the OP $\vec{\phi}_0$
and the  CD $m_0$. The ratio of the kinetic coefficient $\mathring{\Gamma}$ in the relaxation equation for the
OP and the kinetic coefficient $\mathring{\lambda}$ in the  diffusive equation for the CD defines the
dynamical parameter $w$ whose FP value governs the dynamical scaling of the model.

It is worthwhile to summarize some results for model C at a usual
critical point, where the CD can be identified with an energy like
density and the value of specific heat exponent $\alpha$ (in any
case) governs the relevance of the asymmetric static coupling
$\mathring{\gamma}$ between the OP and the CD. Namely, this coupling
is irrelevant in the RG sense - it vanishes at the FP - if the
specific heat exponent is negative, i.e. the specific heat of the
system does not diverge at the critical point. If the specific heat
diverges then there remain two possibilities for the dynamical FP:
either the FP value  of the time scale ratio $w$ between the
timescale of the OP and the CD is  different from zero and finite,
or its FP value is zero or infinite. In the first case strong
dynamical scaling with one timescale for the OP and the CD is
realized with one dynamical scaling exponent $z=2+\alpha/\nu$ ($\nu$
the exponent of the correlation length). In the second case weak
dynamical scaling is present and  the time scale of the OP is
different from the timescale of  the CD, both represented by a
corresponding dynamical critical exponent. This region of the weak
dynamical scaling FP is tiny (see e.g Fig. 1 in Ref.\cite{fomoprl03}).
One should note that at the usual critical point an asymmetric
coupling to the OP as given in Eq. (\ref{hmodc}) is always
'energy-like' independent of its physical origin. That means the
divergence of the CD susceptibility is always described by the
specific heat exponent $\alpha$.

In the case of a multicritical point treated here the situation is
more complicated since there are two OPs and a CD might couple to
both of these OPs. As has been shown in paper I  after a proper
rotation (Eq. (64)) in the OP space  temperature and magnetic like
field directions can be identified. In consequence at the
multicritical point one has to discriminate the case of energy and
magnetization conservation.

The paper is organized as follows. In  section \ref{multC} we extend model C of the $O(n)$ symmetrical critical system to the
case of  a  $O(n_\|)\oplus O(n_\perp)$ symmetrical multicritical point as considered in papers I and II. The renormalization
is performed in section \ref{renorm} and the field theoretic functions are calculated in section \ref{zeta}. Then we
discuss the possible FP and their stability in section \ref{fixp}.  Effective dynamical critical behavior is considered in
section \ref{flows} followed in section \ref{out} by a short summary of the results and an outlook on further work to be
done.

\section{Model C for multicritical points}   \label{multC}

\subsection{Static functional} \label{func}

In order to describe  the
multicritical behavior   the $n$-dimensional space of
the order parameter components is split into two subspaces with dimensions $n_\perp$ and
$n_\|$ with the property $n_\perp+n_\|=n$. The order parameter separates into
\begin{equation}\label{phisplit}
\vec{\phi}_0=\left(\begin{array}{c} \vec{\phi}_{\perp 0} \\  \vec{\phi}_{\| 0}
\end{array}\right)  \ ,
\end{equation}
where $\vec{\phi}_{\perp 0}$ is the $n_\perp$-dimensional order
parameter of the $n_\perp$-subspace, and $\vec{\phi}_{\| 0}$ is the
$n_\|$-dimensional order parameter of the $n_\|$-subspace.
Introducing this separation into the GLW-functional (\ref{hglw}) one
obtains
\begin{eqnarray}\label{hbicrit}
{\cal H}_{Bi}\!=\!\int\! d^dx\Bigg\{\frac{1}{2}\mathring{r}_\perp\vec{\phi}_{\perp 0}
\cdot\vec{\phi}_{\perp 0}+\frac{1}{2}\sum_{i=1}^{n_\perp}\nabla_i\vec{\phi}_{\perp 0}\cdot
\nabla_i\vec{\phi}_{\perp 0}  \nonumber \\
+\frac{1}{2}\mathring{r}_\|\vec{\phi}_{\| 0}
\cdot\vec{\phi}_{\| 0}+\frac{1}{2}\sum_{i=1}^{n_\|}\nabla_i\vec{\phi}_{\| 0}\cdot
\nabla_i\vec{\phi}_{\| 0}  \nonumber \\
+\frac{\mathring{u}_\perp}{4!}\Big(\vec{\phi}_{\perp 0}\cdot\vec{\phi}_{\perp 0}\Big)^2
+\frac{\mathring{u}_\|}{4!}\Big(\vec{\phi}_{\| 0}\cdot\vec{\phi}_{\| 0}\Big)^2
\nonumber \\
+\frac{2\mathring{u}_\times}{4!}\Big(\vec{\phi}_{\perp
0}\cdot\vec{\phi}_{\perp 0}\Big) \Big(\vec{\phi}_{\|
0}\cdot\vec{\phi}_{\| 0}\Big) \Bigg\} \ .
\end{eqnarray}
which represents a multicritical Ginzburg-Landau-Wilson model. The
properties of this functional concerning the renormalization,
regions of stable FPs, and corresponding type of multicritical
behavior has been extensively discussed in paper I (see \cite{phi4}
for earlier references). 
The separation (\ref{phisplit}) has now to be performed in
(\ref{hmodc}). The resulting functional is
\begin{eqnarray}\label{hbicritc}
{\cal H}_{Bi}^{(C)}\!=\!\int\! d^dx\Bigg\{\frac{1}{2}\mathring{\tilde{r}}_\perp\vec{\phi}_{\perp 0}
\cdot\vec{\phi}_{\perp 0}+\frac{1}{2}\sum_{i=1}^{n_\perp}\nabla_i\vec{\phi}_{\perp 0}\cdot
\nabla_i\vec{\phi}_{\perp 0}  \nonumber \\
+\frac{1}{2}\mathring{\tilde{r}}_\|\vec{\phi}_{\| 0}
\cdot\vec{\phi}_{\| 0}+\frac{1}{2}\sum_{i=1}^{n_\|}\nabla_i\vec{\phi}_{\| 0}\cdot
\nabla_i\vec{\phi}_{\| 0}+\frac{1}{2}m_0^2  \nonumber \\
+\frac{\mathring{\tilde{u}}_\perp}{4!}\Big(\vec{\phi}_{\perp 0}\cdot\vec{\phi}_{\perp 0}\Big)^2
+\frac{\mathring{\tilde{u}}_\|}{4!}\Big(\vec{\phi}_{\| 0}\cdot\vec{\phi}_{\| 0}\Big)^2
\nonumber \\
+\frac{2\mathring{\tilde{u}}_\times}{4!}\Big(\vec{\phi}_{\perp
0}\cdot\vec{\phi}_{\perp 0}\Big) \Big(\vec{\phi}_{\|
0}\cdot\vec{\phi}_{\| 0}\Big) \nonumber \\
+\frac{1}{2}\mathring{\gamma_\perp}m_0\vec{\phi}_{\perp 0}\cdot\vec{\phi}_{\perp 0}
+\frac{1}{2}\mathring{\gamma_\|}m_0\vec{\phi}_{\| 0}\cdot\vec{\phi}_{\| 0}
-\mathring{h}m_0\Bigg\} \ .
\end{eqnarray}
Integrating the contributions of the secondary density in the corresponding
partition function, (\ref{hbicritc}) reduces to the static functional (\ref{hbicrit}).
Relations analogous to (\ref{rurel}) between the parameters of the two static
functionals arise. They read
\begin{eqnarray}
\label{rupperprel}
\mathring{r}_\perp=\mathring{\tilde{r}}_\perp+\mathring{\gamma}_\perp\mathring{h}
\ , \qquad  &&
\mathring{u}_\perp=\mathring{\tilde{u}}_\perp-3\mathring{\gamma}_\perp^2  \ ,
\\
\label{rupararel}
\mathring{r}_\|=\mathring{\tilde{r}}_\|+\mathring{\gamma}_\|\mathring{h}
\ , \qquad   &&
\mathring{u}_\|=\mathring{\tilde{u}}_\|-3\mathring{\gamma}_\|^2  \ ,
\\
\label{utimesrel}
&&\mathring{u}_\times=\mathring{\tilde{u}}_\times-3\mathring{\gamma}_\perp
\mathring{\gamma}_\|
\ .
\end{eqnarray}
Because the partition function calculated from (\ref{hbicritc}) is
reducible to a partition function based on (\ref{hbicrit}) by
integration, the correlation functions, or vertex functions
respectively, of the secondary density $m_0$ are exactly related to
correlation functions of the order parameter. This leads to several
relations which are important for the renormalization. In
particular, the average value of $m_0$ and the two-point correlation
function are defined as
\begin{eqnarray}
\label{mavdef}
\langle m_0\rangle\!\!\!&\equiv&\!\!\!\frac{1}{{\cal N}_{Bi}^{(C)}}\int{\cal D}(\phi_{\perp 0},\phi_{\| 0},m_0)
\ m_0\ e^{-{\cal H}_{Bi}^{(C)}} \ , \\
\label{mcorrdef}
\langle m_0\ m_0\rangle\!\!\!&\equiv&\!\!\!\frac{1}{{\cal N}_{Bi}^{(C)}}
\int{\cal D}(\phi_{\perp 0},\phi_{\| 0},m_0)\ m_0\ m_0\ e^{-{\cal H}_{Bi}^{(C)}} \ ,
\nonumber \\
\end{eqnarray}
with
${\cal N}_{Bi}^{(C)}=\int{\cal D}(\phi_{\perp 0},\phi_{\| 0},m_0)\ e^{-{\cal H}_{Bi}^{(C)}}$
as the normalization constant and ${\cal D}(\phi_{\perp 0},\phi_{\| 0},m_0)$ as a suitable
integral measure. Performing the integration over $m_0$ in (\ref{mavdef}) and using
Eqs.(\ref{rupperprel})-(\ref{utimesrel}), the average value of $m_0$ reads
\begin{equation}\label{mav}
\langle m_0\rangle=\mathring{h}
-\mathring{\gamma}_\perp \left\langle \frac{1}{2}\vec{\phi}_{\perp 0}^2\right\rangle
-\mathring{\gamma}_\| \left\langle \frac{1}{2}\vec{\phi}_{\| 0}^2\right\rangle
\end{equation}
where $\vec{\phi}\,^2$ denotes quadratic insertions of the order parameter.
Their average values  on the right hand side of (\ref{mav}),
\begin{eqnarray}
\label{phi2def}
\left\langle \frac{1}{2}\vec{\phi}_{\alpha_{i} 0}^2\right\rangle=\frac{1}{{\cal N}_{Bi}}
\int{\cal D}(\phi_{\perp 0},\phi_{\| 0})
\ \frac{1}{2}\vec{\phi}_{\alpha_{i} 0}^2\ e^{-{\cal H}_{Bi}} \, ,
\end{eqnarray}
are now calculated with the static functional (\ref{hbicrit}) and
${\cal N}_{Bi}=\int{\cal D}(\phi_{\perp 0},\phi_{\| 0})\ e^{-{\cal H}_{Bi}}$. In
order to obtain $\langle m_0\rangle=0$ the conjugated external field is chosen to
\begin{equation}\label{hext}
\mathring{h}=
\mathring{\gamma}_\perp \left\langle \frac{1}{2}\vec{\phi}_{\perp 0}^2\right\rangle
+\mathring{\gamma}_\| \left\langle \frac{1}{2}\vec{\phi}_{\| 0}^2\right\rangle
\end{equation}
Quite analogous by integrating $m_0$ in (\ref{mcorrdef}) one obtains the following relation
for the two-point correlation function of the secondary density
\begin{equation}\label{mmcorr}
\langle m_0\ m_0\rangle_c=1-\mathring{\vec{\gamma}}^T\cdot
\mathring{\mbox{\boldmath$\Gamma$}}^{(0,2)}\cdot \mathring{\vec{\gamma}} \, .
\end{equation}
In (\ref{mmcorr}) where we have introduced
the column matrix
\begin{equation}\label{gammavec}
\mathring{\vec{\gamma}}\equiv\left(\begin{array}{c} \mathring{\gamma}_\perp \\
\mathring{\gamma}_\|
\end{array}\right)  \ .
\end{equation}
The superscript $^T$ indicates a transposed vector or matrix, while the subscript $_c$
on the average at the left hand side of (\ref{mmcorr}) denotes the cummulant
$\langle A\ B\rangle_c\equiv\langle A\ B\rangle-\langle A\rangle\langle B\rangle$.
The matrix
\begin{eqnarray}\label{ga020}
\mathring{\mbox{\boldmath$\Gamma$}}^{(0,2)}&=&\left(\begin{array}{cc}
\mathring{\Gamma}^{(0,2)}_{;\perp\perp} & \mathring{\Gamma}^{(0,2)}_{;\perp\|} \\
\mathring{\Gamma}^{(0,2)}_{;\|\perp} & \mathring{\Gamma}^{(0,2)}_{;\|\|}
\end{array}\right) \nonumber \\
&=&-\left(\begin{array}{cc}
\langle\frac{1}{2}\vec{\phi}_{\perp 0}^2\ \frac{1}{2}\vec{\phi}_{\perp 0}^2\rangle_c &
\langle\frac{1}{2}\vec{\phi}_{\perp 0}^2\ \frac{1}{2}\vec{\phi}_{\| 0}^2\rangle_c \\
\langle\frac{1}{2}\vec{\phi}_{\perp 0}^2\ \frac{1}{2}\vec{\phi}_{\| 0}^2\rangle_c &
\langle\frac{1}{2}\vec{\phi}_{\| 0}^2\ \frac{1}{2}\vec{\phi}_{\| 0}^2\rangle_c
\end{array}\right)
\end{eqnarray}
of two-point vertex functions is related to correlations of
$\phi^2$-insertions. The vertex functions generally have been
introduced in paper I (section III Renormalization), and especially
the matrix (\ref{ga020}) (renormalized counterpart) in Eq.(83)
therein. A third important relation can be obtained by
differentiating the average value (\ref{mavdef}) by $\mathring{h}$
at fixed parameter
$\bigtriangleup\mathring{\tilde{r}}_{\alpha},\mathring{\tilde{u}}_{\alpha},\mathring{\gamma}_{\alpha}$.
In
$\bigtriangleup\mathring{\tilde{r}}_{\alpha}=\mathring{\tilde{r}}_{\alpha}-
\mathring{\tilde{r}}_{\alpha_{i}c}$ the shift of the critical
temperature has been taken into account (for more details see
Appendix A in paper I). As a result one obtains
\begin{equation}\label{dmdh}
\frac{\partial}{\partial\mathring{h}}\ \langle m_0(x)\rangle
\left\vert_{\bigtriangleup\mathring{\tilde{r}}_{\alpha},\mathring{\tilde{u}}_{\alpha},
\mathring{\gamma}_{\alpha}}\right.=
\int dx^\prime\langle m_0(x)\ m_0(x^\prime)\rangle_c   \, .
\end{equation}
Due to relation (\ref{hext}) the external field is function of
$\bigtriangleup\mathring{r}_{\alpha}$. The $h$-derivative in (\ref{dmdh}) can be rewritten
as $\bigtriangleup\mathring{r}_{\alpha}$-derivatives. Finally one obtains
\begin{equation}\label{dmdhend}
\int\!\! dx^\prime\langle m_0(x)\ m_0(x^\prime)\rangle_c=
\mathring{\vec{\gamma}}^T\!\!\!\cdot\frac{\partial}{\partial\bigtriangleup\mathring{\vec{r}}}
\ \langle m_0(x)\rangle
\left\vert_{\bigtriangleup\mathring{\tilde{r}}_{\alpha},\mathring{\tilde{u}}_{\alpha},
\mathring{\gamma}_{\alpha}}\right.
\end{equation}
where we have defined
\begin{equation}\label{ddrvec}
\frac{\partial}{\partial\bigtriangleup\mathring{\vec{r}}}\equiv\left(\begin{array}{c}
\frac{\partial}{\partial\bigtriangleup\mathring{r}_\perp} \\
\frac{\partial}{\partial\bigtriangleup\mathring{r}_\|}
\end{array}\right)  \ .
\end{equation}
All static vertex functions, for the order parameter as well as for the secondary density,
may be calculated with (\ref{hbicritc}) in perturbation expansion as  functions of
the correlation lengths $\{\xi\}\equiv \{\xi_\perp,\xi_\|\}$, the
set of quartic couplings $\{\mathring{\tilde{u}}\}\equiv\{\mathring{\tilde{u}}_\perp,
\mathring{\tilde{u}}_\|,\mathring{\tilde{u}}_\times\}$, the set of asymmetric couplings
$\{\mathring{\gamma}\}\equiv\{\mathring{\gamma}_\perp,\mathring{\gamma}_\|\}$, and the
wave vector modulus $k$. The parameters in the order parameter vertex functions
$\mathring{\bar{\Gamma}}^{(N,L)}_{\alpha_1\cdots\alpha_N;i_1\cdots i_L}$
(for the notation see  Appendix A in paper I) via relations (\ref{rupperprel}) - (\ref{utimesrel}) combine
the corresponding parameters of the multicritical GLW-model (\ref{hbicrit}). Thus all order parameter
vertex functions calculated with (\ref{hbicritc}) have the property
\begin{eqnarray}\label{vertexeq}
\mathring{\bar{\Gamma}}^{(N,L)}_{\alpha_1\cdots\alpha_N;i_1\cdots i_L}
\big(\{\xi\},k,\{\mathring{\tilde{u}}\},\{\mathring{\gamma}\}\big)= \\
\mathring{\Gamma}^{(N,L)}_{\alpha_1\cdots\alpha_N;i_1\cdots i_L}
\big(\{\xi\},k,\{\mathring{u}\}\big) \, .   \nonumber 
\end{eqnarray}
meaning that they are identical to corresponding functions of the
multicritical GLW-model (\ref{hbicrit}). For this reason no
distinction between the correlation lengthes entering the left and
right hand side of (\ref{vertexeq}) is necessary. The correlation
lengths are defined from the  two-point order parameter vertex
functions at the left side with (\ref{hbicritc}), and on the right
side with (\ref{hbicrit}) (see Eqs.(A7) and (A8) in paper I). Vertex
functions of the secondary density can be expressed as functions of
$\{\mathring{u}\}$ instead of $\{\mathring{\tilde{u}}\}$ by using
(\ref{rupperprel})-(\ref{utimesrel}). Especially the two-point
function $\mathring{\bar{\Gamma}}_{mm}=\langle m_0\
m_0\rangle_c^{-1}$, which will be of interest in the following, can
be written as
\begin{equation}\label{gmmbic}
\mathring{\bar{\Gamma}}_{mm}\big(\{\xi\},k,\{\mathring{\tilde{u}}\},\{\mathring{\gamma}\}\big)
=\mathring{\Gamma}_{mm}\big(\{\xi\},k,\{\mathring{u}\},\{\mathring{\gamma}\}\big)
\end{equation}

\subsection{Dynamical model \label{dynmod}}

The dynamical equations of model A in paper II have now to be extended by appending a
diffusion equation for the secondary density.
One obtains
\begin{eqnarray}
\label{dphiperp}
\frac{\partial \vec{\phi}_{\perp 0}}{\partial t}&=&-\mathring{\Gamma}_\perp
\frac{\delta {\mathcal H}_{Bi}^{(C)}}{\delta \vec{\phi}_{\perp 0}}+\vec{\theta}_{\phi_\perp}
\ , \\
\label{dphipar}
\frac{\partial \vec{\phi}_{\|0}}{\partial t}&=&-\mathring{\Gamma}_\|
\frac{\delta {\mathcal H}_{Bi}^{(C)}}{\delta \vec{\phi}_{\|0}}+\vec{\theta}_{\phi_\|}
\, , \\
\label{dmpar}
\frac{\partial m_0}{\partial t}&=&\mathring{\lambda}\nabla^2
\frac{\delta {\mathcal H}_{Bi}^{(C)}}{\delta m_0}+\theta_m \, .
\end{eqnarray}
In addition to the two kinetic coefficients $\mathring{\Gamma}_\perp$ and
$\mathring{\Gamma}_\|$ of the order parameter in the corresponding subspaces, a
kinetic coefficient $\mathring{\lambda}$ of diffusive type for the conserved secondary
density is now present.
The stochastic forces $\vec{\theta}_{\phi_\perp}$, $\vec{\theta}_{\phi_\|}$
and $\theta_m$ fulfill Einstein relations
\begin{eqnarray}
\label{thetaperp}
\langle\theta_{\phi_\perp}^\alpha(x,t)\ \theta_{\phi_\perp}^\beta
(x^\prime,t^\prime)\rangle \!\!\!&=&\!\!\!
2\mathring{\Gamma}_\perp\delta(x-x^\prime)\delta(t-t^\prime)\delta^{\alpha\beta}
\ , \\
\label{thetapara}
\langle\theta_{\phi_\|}^i(x,t)\ \theta_{\phi_\|}^j(x^\prime,t^\prime)
\rangle \!\!\!&=&\!\!\! 2\mathring{\Gamma}_\|\delta(x-x^\prime)\delta(t-t^\prime)
\delta^{ij}
\ , \\
\label{thetam}
\langle\theta_m(x,t)\ \theta_m(x^\prime,t^\prime)
\rangle \!\!\!&=&\!\!\! -2\mathring{\lambda}\nabla^2\delta(x-x^\prime)\delta(t-t^\prime)
\ ,
\end{eqnarray}
with indices $\alpha,\beta=1,\dots , n_\perp$ and $i,j=1,\dots
,n_\|$ corresponding to the two subspaces. The dynamical two-point
vertex function of the secondary density has a general structure
quite analogous to the corresponding functions of the order
parameter (see Eqs.(6) and (7) in paper II). One can write
\begin{eqnarray}\label{gmmstruc}
\mathring{\Gamma}_{m\tilde{m}}\big(\{\xi\},k,\omega\big)=-i\omega
\mathring{\Omega}_{m\tilde{m}}\big(\{\xi\},k,\omega\big)+\mathring{\Gamma}_{mm}\big(\{\xi\},k\big)
\mathring{\lambda}   \nonumber \\
\end{eqnarray}
where $\mathring{\Gamma}_{mm}\big(\{\xi\},k\big)$ is the static
two-point function discussed in the previous subsection and
$\mathring{\Omega}_{m\tilde{m}}\big(\{\xi\},k,\omega\big)$ is a
genuine dynamical function\cite{fomoprl03}. $\tilde{m}$ is the
auxiliary density corresponding to $m$. For shortness we have
dropped the couplings and kinetic coefficients in the argument lists
of (\ref{gmmstruc}).

\section{Renormalization} \label{renorm}

\subsection{Renormalization of the static parameters}  \label{statrenorm}

As a consequence of the discussion at the end of subsection \ref{func} all vertex functions
will be expanded in powers of the quartic couplings $\{\mathring{u}\}$ of the
multicritical GLW-model and the asymmetric couplings $\{\mathring{\gamma}\}$.
The renormalization scheme introduced in section III in paper I remains valid and will be
used in the following. The corresponding definitions and
relations can be found therein and will not be repeated here.
In particular, we implement the minimal subtraction RG
scheme\cite{minsub,review} directly at $d=3$ to the two loop order.
In the current extended model additional renormalizations
for the secondary density $m_0$ and the asymmetric couplings $\{\mathring{\gamma}\}$ have
to be considered. The renormalized counterparts of the secondary density and the
asymmetric couplings are introduced as
\begin{eqnarray}
\label{mren}
m_0&=&Z_mm \\
\label{gammaren}
\mathring{\vec{\gamma}}&=&\kappa^{-\varepsilon/2}Z_m^{-1}
\mbox{\boldmath$Z$}_\phi^{-1}\cdot\mbox{\boldmath$Z$}_\gamma\cdot\vec{\gamma} A_d^{-1/2}
\end{eqnarray}
where $\kappa$ is the usual reference wave vector modulus and $\varepsilon=4-d$. The
geometrical factor $A_d$ and the diagonal matrix $\mbox{\boldmath$Z$}_\phi$ has been defined
in Eq.(8) and Eq.(17) of paper I. With  (\ref{mren}) and (\ref{gammaren}) at hand, the renormalization
for the CD-CD two-point vertex $\Gamma_{mm}$ function readily follows
\begin{equation}\label{gmmren}
\Gamma_{mm}=Z_m^2\mathring{\Gamma}_{mm} \, ,
\end{equation}
The additional $Z$-factor $Z_m$ and the matrix
$\mbox{\boldmath$Z$}_\gamma$ are related to the known
renormalization factors of the multicritical  GLW-model as a
consequence of the reducibility of the extended model to the
multicritical GLW-model.

From the condition that (\ref{dmdhend}) is also valid for the renormalized
counterparts of the appearing quantities the relation
\begin{equation}\label{zgammrel}
\mbox{\boldmath$Z$}_\gamma=Z_m^2\mbox{\boldmath$Z$}_r=
Z_m^2\mbox{\boldmath$Z$}_\phi\cdot\mbox{\boldmath$Z$}_{\phi^2}^T
\end{equation}
follows. For the second equality relation (18) of paper I has been used. The $Z$-factors
of the asymmetric couplings are  determined by the renormalizations of the secondary density
and the $\phi^2$-insertions in the multicritical  GLW-model.

Relation (\ref{mmcorr}) establishes a connection between the
correlation functions of the CD and the $\phi^2$-insertions in the
multicritical GLW-model. This relation should be invariant under
renormalization. Thus the renormalization of the secondary density
is related by
\begin{equation}\label{zmrel}
Z_m^{-2}=1+\vec{\gamma}^T\cdot\mbox{\boldmath$A$}(\{u\})\cdot\vec{\gamma}
\end{equation}
to the additive renormalization $\mbox{\boldmath$A$}(\{u\})$ of the
correlation function of the $\phi^2$-insertions (\ref{ga020}) in the
multicritical GLW-model introduced in Eq.(15) in paper I.

\subsection{Renormalization of the dynamical parameters}

The general form of the renormalization of the auxiliary densities
$\tilde{\phi}_{\perp 0}$ , $\tilde{\phi}_{\| 0}$, and the kinetic coefficients
$\mathring{\Gamma}_\perp$ and $\mathring{\Gamma}_\|$ has been presented within
model A in subsection III A in paper II. It remains valid and will be used in the
following. Of course new contributions occur to the dynamical renormalization
factors especially of the kinetic coefficients
\begin{equation}\label{Gammaren}
\mathring{\Gamma}_\perp=Z_{\Gamma_\perp}\Gamma_{\perp} \ , \qquad
\mathring{\Gamma}_\|=Z_{\Gamma_\|}\Gamma_{\|} \, .
\end{equation}
due to the asymmetric coupling $\vec{\gamma}$.

Within model C additional
renormalizations only are necessary for the auxiliary density
$\tilde{m}_0$ and the kinetic coefficient $\mathring{\lambda}$. Thus we introduce
\begin{equation}\label{mtildelamren}
\tilde{m}_0=Z_{\tilde{m}}\tilde{m} \ , \qquad
\mathring{\lambda}=Z_\lambda\lambda \, .
\end{equation}
In the case of conserved densities the dynamical function
$\mathring{\Omega}_{m\tilde{m}}\big(\{\xi\},k,\omega\big)$ in (\ref{gmmstruc}) does not contain
new dimensional singularities. Therefore the corresponding auxiliary
density $\tilde{m}_0$ needs no independent renormalization. The $Z$-factor $Z_{\tilde{m}}$
is determined by the relation \begin{equation}\label{zmtilde}
Z_{\tilde{m}}=Z_m^{-1}   \, .
\end{equation}
Due to the absence of mode coupling terms the renormalization of the kinetic coefficient
$\lambda$ is completely determined by the static renormalization and $Z_\lambda$ is
\begin{equation}\label{zlambda}
Z_\lambda=Z_m^2  \, .
\end{equation}

\section{$\zeta$ - and $\beta$ - functions} \label{zeta}

As already mentioned in the preceding section the renormalization of the GLW-functional
remains valid. This validates also all $\zeta$- and $\beta$-functions introduced in
section IV in paper I. We do not repeat them here, although they will be used in the
following.

\subsection{Static functions}

Apart from the three $\beta$-functions $\beta_{u_{\perp}}$, $\beta_{u_{\times}}$ and
$\beta_{u_{\|}}$, and the two $\zeta$-matrices $\mbox{\boldmath$B$}_{\phi^2}$
and $\mbox{\boldmath$\zeta$}_{\phi^2}$ appearing in the multicritical
GLW-model (see section IV in paper I), an additional $\zeta$-function $\zeta_m$ and
a column matrix of $\beta$-functions for the asymmetric coupling (\ref{gammavec}) have to be introduced.
The relations between the renormalization factors discussed in subsection
\ref{statrenorm}
give rise to corresponding relations between the $\zeta$- and $\beta$-functions.
It follows immediately from (\ref{zmrel})
\begin{equation}\label{zetam}
\zeta_m(\{u\},\{\gamma\})\equiv\frac{d\ln Z_m^{-1}}{d\ln\kappa}=\frac{1}{2}
\vec{\gamma}^T\cdot\mbox{\boldmath$B$}_{\phi^{2}}(\{u\})\cdot\vec{\gamma} \, ,
\end{equation}
where $\mbox{\boldmath$B$}_{\phi^{2}}(\{u\})$ has been defined in Eq.(30) in paper I.
The $\kappa$-derivatives, also in the following definitions, always are taken at
fixed unrenormalized parameters. Inserting the two loop expression of
$\mbox{\boldmath$B$}_{\phi^{2}}(\{u\})$ (see Eq.(31) in paper I) we obtain
\begin{equation}\label{zetam2l}
\zeta_m(\{u\},\{\gamma\})=\frac{n_\perp}{4}\ \gamma_\perp^2+\frac{n_\|}{4}\ \gamma_\|^2  \, .
\end{equation}
The column matrix of the  $\beta$-functions for the asymmetric
coupling $\vec{\gamma}$ is defined as
\begin{equation}\label{betagamdef}
\vec{\beta}_\gamma(\{u\},\{\gamma\})\equiv\kappa\frac{d\vec{\gamma}}{d\kappa}
\end{equation}
Inserting Eq.(\ref{gammaren}) into the above definition one obtains together with
relation (\ref{zgammrel}) the expression
\begin{equation}\label{betagam}
\vec{\beta}_\gamma(\{u\},\{\gamma\})=\left[\left(-\frac{\varepsilon}{2}+\zeta_m\right)
\mbox{\boldmath$1$}+\mbox{\boldmath$\zeta$}_{\phi^2}^T(\{u\})\right]\cdot
\vec{\gamma}  \ .
\end{equation}
There \mbox{\boldmath$1$} denotes the two dimensional unit matrix. The matrix
$\mbox{\boldmath$\zeta$}_{\phi^2}(\{u\})$ has been introduced in paper I (see Eq.(22)).
The $\zeta$-function $\zeta_m$ is exactly known from (\ref{zetam}). Thus finally we arrive
at
\begin{eqnarray}\label{betagamend}
\vec{\beta}_\gamma(\{u\},\{\gamma\})=\Bigg[\left(-\frac{\varepsilon}{2}
+\frac{1}{2}\vec{\gamma}^T\cdot\mbox{\boldmath$B$}_{\phi^{2}}(\{u\})\cdot
\vec{\gamma}\right)\mbox{\boldmath$1$}  \nonumber \\
+\mbox{\boldmath$\zeta$}_{\phi^2}^T(\{u\})\Bigg]\cdot
\vec{\gamma}  \ .
\end{eqnarray}
The above expression is valid in all orders of perturbation expansion.
$\mbox{\boldmath$B$}_{\phi^2}(\{u\})$ and $\mbox{\boldmath$\zeta$}_{\phi^2}(\{u\})$
are calculated in loop expansion within the multicritical GLW-model. Their two loop
expressions have been given in Eq.(31) and (23)-(26) in paper I.

\subsection{Dynamical functions}

Using relation (\ref{zlambda}) the $\zeta$-function $\zeta_\lambda$ corresponding
to the kinetic coefficient $\lambda$ is simply given by
\begin{equation}\label{zetalambda}
\zeta_\lambda(\{u\},\{\gamma\})\equiv\frac{d\ln Z_\lambda^{-1}}{d\ln\kappa}=
2\zeta_m(\{u\},\{\gamma\})
\end{equation}
The dynamical $\zeta$-functions of the kinetic coefficients of the order parameter
are defined by
\begin{equation}\label{zetaop}
\zeta_{\Gamma_\alpha}^{(C)}(\{u\},\{\gamma\},\{w\})\equiv\frac{d\ln Z_{\Gamma_\alpha}^{-1}}{d\ln\kappa}
\, , \quad  \alpha=\|,\perp \, .
\end{equation}
In the model C dynamics, they get non-trivial contributions from the
asymmetric couplings $\gamma_\perp$ and $\gamma_\|$. They read now
in two loop order
\begin{eqnarray}
\label{zetagaperp2l}
\zeta_{\Gamma_\perp}^{(C)}\big(\{u\},\{\gamma\},\{w\})\big)=
\bar{\zeta}^{(C_\perp)}\big(u_\perp,\gamma_\perp,w_\perp\big)\nonumber \\
-\frac{n_\|}{4}\frac{w_\perp\gamma_\perp\gamma_\|}{1+w_\perp}\Bigg[
\frac{2}{3}u_\times+\frac{w_\perp\gamma_\perp\gamma_\|}{1+w_\perp}\Bigg]
\Bigg(1+\ln\frac{2v}{1+v} \nonumber \\
-\Big(1+\frac{2}{v}\Big)\ln\frac{2(1+v)}{2+v}\Bigg)+
\zeta_{\Gamma_\perp}^{(A)}\big(u_\perp,u_\times,v\big) \, , \nonumber \\
\end{eqnarray}
\begin{eqnarray}
\label{zetagapara2l}
\zeta_{\Gamma_\|}^{(C)}\big(\{u\},\{\gamma\},\{w\})\big)=
\bar{\zeta}^{(C_\|)}\big(u_\|,\gamma_\|,w_\|\big)\nonumber \\
-\frac{n_\perp}{4}\frac{w_\|\gamma_\|\gamma_\perp}{1+w_\|}\Bigg[
\frac{2}{3}u_\times+\frac{w_\|\gamma_\|\gamma_\perp}{1+w_\|}\Bigg]
\Bigg(1+\ln\frac{2}{1+v} \nonumber \\
-(1+2v)\ln\frac{2(1+v)}{1+2v}\Bigg)+
\zeta_{\Gamma_\|}^{(A)}\big(u_\|,u_\times,v\big) \nonumber \\
\end{eqnarray}
where we have defined the timescale ratios
\begin{equation}\label{wperppara}
w_\perp=\frac{\Gamma_\perp}{\lambda} \ , \qquad w_\|=\frac{\Gamma_\|}{\lambda}
\end{equation}
The ratio $v$ is equally defined to paper II as the ratio
\begin{equation}\label{vdef}
v\equiv\frac{\Gamma_\|}{\Gamma_\perp}=\frac{w_\|}{w_\perp} \ ,
\end{equation}
and is therefore a function of $w_\perp$ and $w_\|$. In
(\ref{zetagaperp2l}) and (\ref{zetagapara2l}) several
$\zeta$-functions of known subsystems already has been introduced.
$\zeta_{\Gamma_{\alpha}}^{(A)}\big(u_{\alpha},u_\times,v\big)$ with
$\alpha=\|$ or $\perp$ are the $\zeta$-functions of the full
multicritical model A presented explicitly in paper II (see Eqs (14)
and (15) therein),
\begin{eqnarray}
\label{zetagperp}
\zeta_{\Gamma_\perp}^{(A)}&=&\frac{n_\perp+2}{36}\ u_\perp^2\left(3\ln\frac{4}{3}
-\frac{1}{2}\right) \\
&+&\frac{n_\|}{36}\ u_\times^2\left[\frac{2}{v}\ln\frac{2(1+v)}{2+v}
+\ln\frac{(1+v)^2}{v(2+v)}-\frac{1}{2}\right] \, , \nonumber\\
\label{zetagpara}
\zeta_{\Gamma_\|}^{(A)}&=&\frac{n_\|+2}{36}\ u_\|^2\left(3\ln\frac{4}{3}
-\frac{1}{2}\right)   \\
&+&\frac{n_\perp}{36}\ u_\times^2\left[2v\ln\frac{2(1+v)}{1+2v}
+\ln\frac{(1+v)^2}{1+2v}-\frac{1}{2}\right] \, , \nonumber
\end{eqnarray}
where $\bar{\zeta}^{(C_\alpha)}\big(u,\gamma,w\big)$ are the genuine
$\zeta$-functions of model C within
the $n_{\alpha}$-component subspaces without pure
fourth order coupling terms (pure model A terms). They have been
given explicitly in \cite{fomo04} for a $n$-component system in two
loop order. These contributions of model C in the
$n$-component subspaces without the corresponding model A
terms are\cite{fomoprl03}
\begin{eqnarray}
\label{modelcsub}
\bar{\zeta}^{(C_\alpha)}\big(u,\gamma,
w\big)=\frac{w\gamma^2}{1+w}\Bigg\{1 \nonumber \\
-\frac{1}{2}\Bigg[\frac{n_\alpha+2}{3}u\Big(1-3\ln\frac{4}{3}\Big)
+\frac{w\gamma^2}{1+w}\Bigg(\frac{n_{\alpha}}{2}
\nonumber \\
-\frac{w}{1+w}-\frac{3(n_{\alpha}+2)}{2}\ln\frac{4}{3}\nonumber \\
-\frac{(1+2w)}{1+w}\ln\frac{(1+w)^2}{1+2w}
\Bigg)\Bigg]\Bigg\} \, .
 \nonumber \\
\end{eqnarray}
The $\beta$-functions corresponding to the timescale ratios (\ref{wperppara})
and (\ref{vdef}) can be expressed in terms of the corresponding $\zeta$-functions of the kinetic coefficients:
\begin{eqnarray}
\label{betav}
\beta_v&\equiv& \kappa\frac{dv}{d\kappa}=v\big(\zeta_{\Gamma_\|}^{(C)}
-\zeta_{\Gamma_\perp}^{(C)}\big) \, ,
 \\
\label{betawpara}
\beta_{w_\|}&\equiv& \kappa\frac{dw_\|}{d\kappa}=w_\|\big(\zeta_{\Gamma_\|}^{(C)}
-\zeta_\lambda\big) \, ,
 \\
\label{betawperp}
\beta_{w_\perp}&\equiv& \kappa\frac{dw_\perp}{d\kappa}=w_\perp\big(\zeta_{\Gamma_\perp}^{(C)}
-\zeta_\lambda\big)  \, ,
\end{eqnarray}
with the $\kappa$-derivatives taken at fixed unrenormalized parameters.

Note that these equations are not independent but one of the three equations can be eliminated
by the relation  $v(l)=w_\|(l)/w_\perp(l)$, which of course holds also for the initial conditions.

\section{Fixed points and their stability} \label{fixp}

\subsection{Static fixed points}

The FPs of the couplings $u_a$ and their stability have been studied
in section V of paper I. Their values and the corresponding
transient exponents have been listed in Table I there. Let us
recall that depending on the values of $n_\|$ and $n_\perp$ one of
the following FPs is stable and governs multicritical behavior:
 the isotropic Heisenberg FP ${\cal H}(n_\|+n_\perp)$ with
$u^\star_\|=u^\star_\perp=u_\times^\star$, the decoupling FP ${\cal
D}$ with $u^\star_\|\neq 0$, $u^\star_\perp\neq 0$ and
$u^\star_\times= 0$, and the biconical FP with $u^\star_\|\neq 0$,
$u^\star_\perp\neq 0$, and $u^\star_\times\neq 0$. For each of the
FPs of paper I one can now determine the FP values for
$\gamma_\perp$ and $\gamma_\|$ from the Eq.
\begin{eqnarray}\label{fpgamma}
&&\vec{\beta}_{\gamma}(\{u^\star\},\{\gamma^\star\})= 
\Bigg[\left(-\frac{\varepsilon}{2}
+\frac{1}{2}\vec{\gamma}^{\star T}\cdot\mbox{\boldmath$B$}_{\phi^{2}}(\{u^\star\})\cdot
\vec{\gamma}^\star\right)\mbox{\boldmath$1$}  \nonumber \\  
&&+\mbox{\boldmath$\zeta$}_{\phi^2}^T(\{u^\star\})\Bigg]\cdot
\vec{\gamma}^\star=0 \, .
\end{eqnarray}
This splits each FP of paper I into a set of FPs in the combined $\{u\}$-$\{\gamma\}$-space,
which are equivalent in statics but different in dynamics.
Given the formula (\ref{betagamend}) for $\vec{\beta}_\gamma$,
one can see, that Eq. (\ref{fpgamma}) includes two equations which
have to be solved for given FP values of the quartic couplings
$\{u^\star\}$. The static FPs of paper I can be roughly separated
into three classes: i) the Gaussian FP ${\mathcal G}$, with
$u_a^\star=0$ for all couplings; ii) the decoupling FPs
$\mathcal{H}(n_\perp),\mathcal{H}(n_\|), \mathcal{D}$ where
$u_\times^\star=0$; iii) the isotropic Heisenberg and biconical FPs
$\mathcal{H}(n_\perp+n_\|), \mathcal{B}$ where all $u_a^\star$ are
different from zero. Henceforth we list the FPs values of the
corresponding asymmetric couplings $\gamma^\star_\|$ and
$\gamma^\star_\perp$ in Tab. \ref{tab1}, that summarizes our
analysis given below. Note, that $\gamma_\perp^\star=0$,
$\gamma_\|^\star=0$ is of course always a solution of equation
(\ref{fpgamma}), independent which values $\{u^\star\}$ have. We do
not list this trivial solution explicitly in Table \ref{tab1},
although this may be the stable FP for definite values of $n_\perp$
and $n_\|$ in some cases.

\subsubsection{Gaussian fixed point ${\mathcal G}$}
 At this FP  one has $\mbox{\boldmath$\zeta$}_{\phi^2}=
\mbox{\boldmath$0$}$ and the two equations in (\ref{fpgamma}) reduce to the
condition
\begin{equation}\label{fpgauss}
\frac{n_\perp}{2}\ \gamma_\perp^{\star 2}+\frac{n_\|}{2}\gamma_\|^{\star 2}
=\varepsilon \ ,
\end{equation}
which is valid in all orders of perturbation expansion. The above equation defines
a line of FPs.

\subsubsection{Heisenberg and decoupling fixed points $\mathcal{H}(n_\perp),\mathcal{H}(n_\|), \mathcal{D}$ }
 At these FPs, where the cross coupling $u_\times$ vanishes, the matrix
$\mbox{\boldmath$\zeta$}_{\phi^2}$ has the form
\begin{eqnarray}\label{zetap2dec}
\mbox{\boldmath$\zeta$}_{\phi^2}(u_\times=0)=\left(\begin{array}{cc}
\zeta_{\phi^2}^{(n_\perp)}(u_\perp) & 0 \\  0 & \zeta_{\phi^2}^{(n_\|)}(u_\|)
\end{array}\right)  \, .
\end{eqnarray}
The function $\zeta_{\phi^2}^{(n_{\alpha})}(u_{\alpha})$ is the well known
$\zeta$-function of the $n_{\alpha}$-component isotropic system.
Eq.(\ref{fpgamma}) reduces to
\begin{eqnarray}
\label{fpdecouple1}
\left(-\frac{\varepsilon}{2}
+\frac{1}{2}\vec{\gamma}^{\star T}\cdot\mbox{\boldmath$B$}_{\phi^{2}}(\{u^\star\})\cdot
\vec{\gamma}^\star +\zeta_{\phi^2}^{(n_\perp)}(u_\perp^\star)
\right)\gamma_\perp=0 \, , \nonumber \\ \\
\label{fpdecouple2}
\left(-\frac{\varepsilon}{2}
+\frac{1}{2}\vec{\gamma}^{\star T}\cdot\mbox{\boldmath$B$}_{\phi^{2}}(\{u^\star\})\cdot
\vec{\gamma}^\star +\zeta_{\phi^2}^{(n_\|)}(u_\|^\star)
\right)\gamma_\|=0 \, , \nonumber \\
\end{eqnarray}
where the matrix $\mbox{\boldmath$B$}_{\phi^2}$ is of the form
\begin{eqnarray}\label{bp2dec}
\mbox{\boldmath$B$}_{\phi^2}(u_\times=0)=\left(\begin{array}{cc}
B_{\phi^2}^{(n_\perp)}(u_\perp) & 0 \\  0 & B_{\phi^2}^{(n_\|)}(u_\|)
\end{array}\right)   \, .
\end{eqnarray}
The non trivial FP values for $\gamma_\perp$ and $\gamma_\|$ resulting
from Eqs.(\ref{fpdecouple1}) and (\ref{fpdecouple2}) are listed in Table \ref{tab1}.
They are valid in all orders of perturbation expansion.

We want to remark that in Table \ref{tab1} only FP which exist for arbitrary
order parameter component numbers are given. For special $n$-values a line of
FPs exist where both asymmetric couplings
$\gamma_{\alpha}$ are different from zero. In the case $n_\perp=n_\|=n/2$ one
has $u_\perp^\star=u_\|^\star=\bar{u}^\star$ and the
$\zeta$-functions in (\ref{fpdecouple2}) and (\ref{fpdecouple2}) are equal
leading to $\gamma_\perp^{\star 2}+\gamma_\|^{\star 2}=\big(\varepsilon
-2\zeta_{\phi^2}^{(n/2)}(\bar{u}^\star)\big)/B_{\phi^2}^{(n/2)}(\bar{u}^\star)$.

\subsubsection{Isotropic Heisenberg and biconical FPs $\mathcal{H}(n_\perp+n_\|), \mathcal{B}$ }
At the isotropic Heisenberg FP $\mathcal{H}(n_\perp+n_\|)$ and the
biconical FP $\mathcal{B}$, where all couplings $u_a$ are different
from zero, it is more convenient to transform the matrix
$\mbox{\boldmath$\zeta$}_{\phi^2}$ into its diagonal form with the transformation
\begin{equation}\label{zetatransform}
\left(\begin{array}{cc} \zeta_+ & 0 \\ 0 & \zeta_- \end{array}\right)
=\mbox{\boldmath$P$}^{-1}\cdot\mbox{\boldmath$\zeta$}_{\phi^2}^T
\cdot\mbox{\boldmath$P$}
\end{equation}
introduced in section VI.B in paper I. Inserting (\ref{zetatransform}) into Eq.(\ref{fpgamma}) leads to
the transformed $\beta$-function
\begin{equation}\label{betatransform}
\vec{\beta}_\gamma=\mbox{\boldmath$P$}\cdot\vec{\beta}_{\gamma_\pm}
\end{equation}
with
\begin{eqnarray}\label{betagammapm}
\vec{\beta}_{\gamma_\pm}=\Bigg[\left(-\frac{\varepsilon}{2}
+\frac{1}{2}\vec{\gamma}^T\cdot\mbox{\boldmath$B$}_{\phi^{2}}\cdot
\vec{\gamma}\right)\mbox{\boldmath$1$}
+\left(\begin{array}{cc} \zeta_+ & 0 \\ 0 & \zeta_- \end{array}\right) \Bigg]\cdot
\vec{\gamma}_\pm  \ . \nonumber \\
\end{eqnarray}
Here, the transformed asymmetric coupling column matrix is defined
as
\begin{equation}\label{gammatransform}
\vec{\gamma}_\pm\equiv \left(\begin{array}{c} \gamma_+ \\
\gamma_- \end{array}\right)
=\mbox{\boldmath$P$}^{-1}\cdot\vec{\gamma} \, .
\end{equation}
Note, that the scalar quantity
\begin{equation}\label{scalar}
\vec{\gamma}^T\cdot\mbox{\boldmath$B$}_{\phi^{2}}\cdot\vec{\gamma}=
\vec{\gamma}_\pm^T\cdot\mbox{\boldmath$B$}_{\phi^2}^{(\pm)}\cdot\vec{\gamma}_\pm
\ ,
\end{equation}
where $\mbox{\boldmath$B$}_{\phi^2}^{(\pm)}=\mbox{\boldmath$P$}^T\cdot
\mbox{\boldmath$B$}_{\phi^2}\cdot\mbox{\boldmath$P$}$, is invariant under transformation.
Therefore in (\ref{betagammapm}) it is written in the untransformed form. At the FP
Eq.(\ref{betatransform}) reduces to the condition
\begin{equation}\label{betagpmtr}
\vec{\beta}_{\gamma_\pm}(\{u^\star\},\{\gamma^\star\})=0
\end{equation}
because the determinant of matrix {\boldmath$P$} does not vanish.
Subsequently, Eq. (\ref{betagammapm}) leads to two FP equations
\begin{eqnarray}
\label{fixpmp}
\Big[-\varepsilon
+\vec{\gamma}^{\star T}\cdot\mbox{\boldmath$B$}_{\phi^{2}}^\star\cdot
\vec{\gamma}^\star+2\zeta_+^\star \Big]\gamma_+^\star  =0 \\
\label{fixpmm}
\Big[-\varepsilon
+\vec{\gamma}^{\star T}\cdot\mbox{\boldmath$B$}_{\phi^{2}}^\star\cdot
\vec{\gamma}^\star+2\zeta_-^\star \Big]\gamma_-^\star  =0
\end{eqnarray}
In the above equations we have introduced the short hand notations
$\mbox{\boldmath$B$}_{\phi^{2}}^\star\equiv\mbox{\boldmath$B$}_{\phi^{2}}(\{u^\star\})$
and $\zeta_\pm^\star\equiv\zeta_\pm(\{u^\star\})$.
If both transformed asymmetric couplings $\gamma_+^\star$ and $\gamma_-^\star$
are different from zero, the above two equations lead to the condition
$\zeta_+^\star=\zeta_-^\star$.  This condition is not valid if {\it all} quartic couplings
$u_a^\star$ are different from zero. Thus at least one of the two transformed
asymmetric couplings, $\gamma_+$ or $\gamma_-$, has to be zero at the FP.
The transformation matrix {\boldmath$P$} has
been presented in Eq.(64) in paper I. Expressed in terms of the $\zeta$-functions it reads
\begin{equation}\label{pzeta}
\mbox{\boldmath$P$}=\left(\begin{array}{cc} P_{11} & P_{12} \\ P_{21} & P_{22}
\end{array}\right)
=\left(\begin{array}{cc} 1 &
\frac{\big[\mbox{\boldmath$\zeta$}_{\phi^2}\big]_{21}}
{\zeta_--\big[\mbox{\boldmath$\zeta$}_{\phi^2}\big]_{11}} \\
\frac{\big[\mbox{\boldmath$\zeta$}_{\phi^2}\big]_{12}}
{\zeta_+-\big[\mbox{\boldmath$\zeta$}_{\phi^2}\big]_{22}} & 1
\end{array}\right)
\end{equation}
where $[\mbox{\boldmath$\zeta$}_{\phi^2}]_{ij}$ are the elements of the matrix
$\mbox{\boldmath$\zeta$}_{\phi^2}$ (for the two loop expressions see Eqs.(23)-(26)
in paper I).
\begin{widetext}

\begin{table*}[h]
\centering \tabcolsep=5mm
\begin{tabular}{llll}
 \hline \hline
   FP &  $\gamma_\perp^{\star}/\gamma_\|^{\star}$ & $\gamma_\perp^{\star 2}$
 & $\gamma_\|^{\star 2}$
   \\ \hline \hline
   ${\mathcal G}$ &  line of FPs (\ref{fpgauss}) & line of FPs (\ref{fpgauss}) & line of FPs (\ref{fpgauss})
   \\  \hline
& 0 & 0 & $\frac{2}{n_\|}\varepsilon$  \\
\raisebox{2.5ex}[-1.5ex]{$\mathcal{H}(n_\perp)$} &   $\infty$
& $\frac{\varepsilon-2\zeta_{\phi^2}^{(n_\perp)}(u_\perp^\star)}
    {B_{\phi^2}^{(n_\perp)}(u_\perp^\star)}$  & 0
    \\ \hline
& $\infty$ &  $\frac{2}{n_\perp}\varepsilon$ & 0   \\
\raisebox{2.5ex}[-1.5ex]{$\mathcal{H}(n_\|)$} & 0
& 0 & $\frac{\varepsilon-2\zeta_{\phi^2}^{(n_\|)}(u_\|^\star)}
    {B_{\phi^2}^{(n_\|)}(u_\|^\star)}$
    \\ \hline
& $\infty$ &  $\frac{\varepsilon-2\zeta_{\phi^2}^{(n_\perp)}(u_\perp^\star)}
    {B_{\phi^2}^{(n_\perp)}(u_\perp^\star)}$  & 0   \\
\raisebox{2.5ex}[-1.5ex]{$\mathcal{D}$} & 0
& 0  & $\frac{\varepsilon-2\zeta_{\phi^2}^{(n_\|)}(u_\|^\star)}
    {B_{\phi^2}^{(n_\|)}(u_\|^\star)}$
    \\  \hline \\
      & $\frac{\zeta_+^\star-\big[\mbox{\boldmath$\zeta$}_{\phi^2}^\star\big]_{22}}
{\big[\mbox{\boldmath$\zeta$}_{\phi^2}^\star\big]_{12}}$
  &  $\frac{2(\varepsilon-2\zeta_+^\star)}{n_\perp
+n_\|\left(\frac{\big[\mbox{\boldmath$\zeta$}_{\phi^2}^\star\big]_{12}}
{\zeta_+^\star-\big[\mbox{\boldmath$\zeta$}_{\phi^2}^\star\big]_{22}}\right)^2}$
&  $\frac{2(\varepsilon-2\zeta_+^\star)}{
n_\perp\left(\frac{\zeta_+^\star-\big[\mbox{\boldmath$\zeta$}_{\phi^2}^\star\big]_{22}}
{\big[\mbox{\boldmath$\zeta$}_{\phi^2}^\star\big]_{12}}\right)^2
+n_\|}$ \\
 \raisebox{5.0ex}[-1.5ex]{$\mathcal{H}(n_\perp+n_\|),\ \mathcal{B}$} &
$\frac{\big[\mbox{\boldmath$\zeta$}_{\phi^2}^\star\big]_{21}}
{\zeta_-^\star-\big[\mbox{\boldmath$\zeta$}_{\phi^2}^\star\big]_{11}}$
&  $\frac{2(\varepsilon-2\zeta_-^\star)}{n_\perp+
n_\|\left(\frac{\zeta_-^\star-\big[\mbox{\boldmath$\zeta$}_{\phi^2}^\star\big]_{11}}
{\big[\mbox{\boldmath$\zeta$}_{\phi^2}^\star\big]_{21}}\right)^2}$ &
$\frac{2(\varepsilon-2\zeta_-^\star)}{
n_\perp\left(\frac{\big[\mbox{\boldmath$\zeta$}_{\phi^2}^\star\big]_{21}}
{\zeta_-^\star-\big[\mbox{\boldmath$\zeta$}_{\phi^2}^\star\big]_{11}}\right)^2
+n_\|}$ \\ \\
 \hline \hline
\end{tabular}
\caption{Fixed points of the asymmetric couplings $\gamma_\perp$ and $\gamma_\|$
of the extended $O(n_\|)\oplus O(n_\perp)$ model. The values of the isotropic
Heisenberg FP $\mathcal{H}(n_\perp+n_\|)$ and the Biconical FP $\mathcal{B}$ are valid
in two loop order because Eq.(\ref{scalar2l}) has been used. For all other FPs
the expressions are valid in all orders of perturbation expansion. \label{tab1}}
\end{table*}
\end{widetext}   

Let us now consider the two cases where one of the asymmetric couplings is nonzero.\\
{\bf Case a:} $\gamma_+^\star\ne 0$, $\gamma_-^\star=0$:
Taking into account Eq. (\ref{gammatransform})  the condition for a vanishing $\gamma_-^\star$
reads $-P_{21}\gamma_\perp^\star+P_{11}\gamma_\|^\star=0$. Given the matrix elements (\ref{pzeta})
it can be rewritten as
\begin{equation}\label{vgpagpe}
\frac{\gamma_\|^\star}{\gamma_\perp^\star}=
\frac{\big[\mbox{\boldmath$\zeta$}_{\phi^2}^\star\big]_{12}}
{\zeta_+^\star-\big[\mbox{\boldmath$\zeta$}_{\phi^2}^\star\big]_{22}}
\end{equation}
At finite $\gamma_+^\star$ the bracket in Eq.(\ref{fixpmp}) has to vanish, which
results in the condition
\begin{equation}\label{gplfix}
2\zeta^\star_m=\vec{\gamma}^{\star T}\cdot\mbox{\boldmath$B$}_{\phi^{2}}^\star\cdot
\vec{\gamma}^\star=\varepsilon-2\zeta_+^\star =\frac{\alpha}{\nu} \ .
\end{equation}
The last equality uses the definition of the asymptotic exponents derived in paper I (Eq. (90) there).\\
{\bf Case b:} $\gamma_+^\star= 0$, $\gamma_-^\star\ne 0$: In this case Eq.
(\ref{gammatransform}) leads immediately to the condition
$P_{22}\gamma_\perp^\star-P_{12}\gamma_\|^\star=0$. Inserting (\ref{pzeta})
gives
\begin{equation}\label{vgpegpa}
\frac{\gamma_\perp^\star}{\gamma_\|^\star}=
\frac{\big[\mbox{\boldmath$\zeta$}_{\phi^2}^\star\big]_{21}}
{\zeta_-^\star-\big[\mbox{\boldmath$\zeta$}_{\phi^2}^\star\big]_{11}} \, .
\end{equation}
At finite $\gamma_-^\star$ the bracket in Eq.(\ref{fixpmm}) has to vanish, which
results in the condition
\begin{equation}\label{gmifix}
2\zeta^\star_m=\vec{\gamma}^{\star T}\cdot\mbox{\boldmath$B$}_{\phi^{2}}^\star\cdot
\vec{\gamma}^\star=\varepsilon-2\zeta_-^\star=2\frac{\phi}{\nu}-d \ .
\end{equation}
Again, the last equality uses the definition of the asymptotic exponents derived in paper I (Eq. (82) there).
The above equations (\ref{vgpagpe}), (\ref{gplfix}) and (\ref{vgpegpa}),
(\ref{gmifix}) respectively, determine the FP values of the two asymmetric
couplings in the corresponding cases. The relations are valid in all orders of
perturbation expansion.

Since in two loop order $\mbox{\boldmath$B$}_{\phi^{2}}^\star$ is diagonal and
independent of the couplings $\{u\}$ (see Eq.(31) in paper I) the
left hand sides of Eqs (\ref{gplfix}) and (\ref{gmifix}) read
\begin{equation}\label{scalar2l}
\vec{\gamma}^{\star T}\cdot\mbox{\boldmath$B$}_{\phi^{2}}^\star\cdot \vec{\gamma}^\star=
\gamma_\perp^{\star 2}\frac{n_\perp}{2}+\gamma_\|^{\star 2}\frac{n_\|}{2}   \, .
\end{equation}
In consequence the asymmetric static couplings are zero when the
exponent expressions  on the right hand side of (\ref{gplfix}) and (\ref{gmifix}) are zero. This is the
case if the specific heat like CD   and/or the magnetic like CD
susceptibility do not diverge.

Using (\ref{scalar2l}) together with (\ref{vgpagpe})-(\ref{gmifix}) leads to the FP values of the asymmetric
couplings $\gamma_\|^{\star 2}$ and $\gamma_\perp^{\star 2}$\\
{\bf Case a:} $\gamma_+^\star\ne 0$, $\gamma_-^\star=0$:
\begin{equation}\label{gapest1}
\gamma_\perp^{\star 2}=\frac{2(\varepsilon-2\zeta_+^\star)}{n_\perp
+n_\|\left(\frac{\big[\mbox{\boldmath$\zeta$}_{\phi^2}^\star\big]_{12}}
{\zeta_+^\star-\big[\mbox{\boldmath$\zeta$}_{\phi^2}^\star\big]_{22}}\right)^2}  \, ,
\end{equation}
\begin{equation}\label{gapast1}
\gamma_\|^{\star 2}=\frac{2(\varepsilon-2\zeta_+^\star)}{
n_\perp\left(\frac{\zeta_+^\star-\big[\mbox{\boldmath$\zeta$}_{\phi^2}^\star\big]_{22}}
{\big[\mbox{\boldmath$\zeta$}_{\phi^2}^\star\big]_{12}}\right)^2
+n_\|}  \, .
\end{equation}
{\bf Case b:} $\gamma_+^\star=0$, $\gamma_-^\star\ne 0$:
\begin{equation}\label{gapest2}
\gamma_\perp^{\star 2}=\frac{2(\varepsilon-2\zeta_-^\star)}{n_\perp+
n_\|\left(\frac{\zeta_-^\star-\big[\mbox{\boldmath$\zeta$}_{\phi^2}^\star\big]_{11}}
{\big[\mbox{\boldmath$\zeta$}_{\phi^2}^\star\big]_{21}}\right)^2
} \, ,
\end{equation}
\begin{equation}\label{gapast2}
\gamma_\|^{\star 2}=\frac{2(\varepsilon-2\zeta_-^\star)}{
n_\perp\left(\frac{\big[\mbox{\boldmath$\zeta$}_{\phi^2}^\star\big]_{21}}
{\zeta_-^\star-\big[\mbox{\boldmath$\zeta$}_{\phi^2}^\star\big]_{11}}\right)^2
+n_\|}   \, .
\end{equation}
Note that the ratios in (\ref{vgpagpe})  and (\ref{vgpegpa}) might be negative leading to a negative product
$\gamma^\star_\|\gamma^\star_\perp$.

\begin{figure}[t,b]
      \centering{
       \epsfig{file=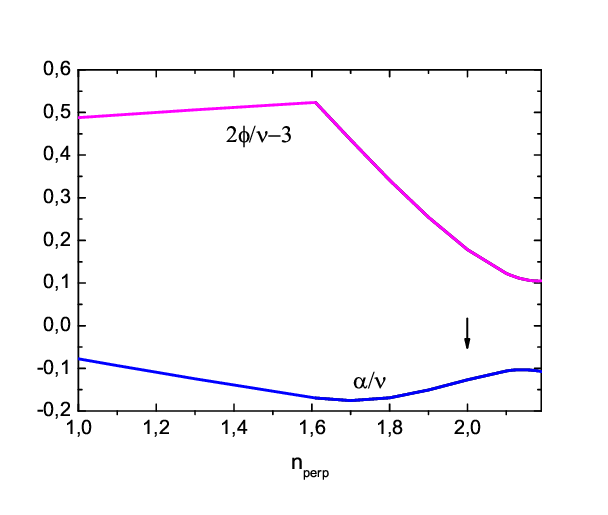,width=9cm,angle=0}  }
     \caption{ \label{exponents} Exponents $\alpha/\nu$ and $2\phi/\nu-3$ appearing in Eqs. (\ref{gplfix}) and
(\ref{gmifix}) at $n_\|=1$ as function of $n_\perp$. Note that via relation (\ref{zstrong}) $2\phi/\nu-3=z-2$ the strong
dynamical scaling exponent $z$ results.}
   \end{figure}
The explicit values of the above FPs depend on wether the isotropic
Heisenberg or the biconical FP is inserted into the
$\zeta$-functions. Eqs.(\ref{gapest1})-(\ref{gapast2}) are valid up
to two loop order. In three loop order it is known from the
isotropic GLW-model that the function $B_{\phi^2}$ gets
$u^2$-contributions \cite{changhoughton80}. In the multicritical
GLW-model the matrix $\mbox{\boldmath$B$}_{\phi^{2}}$ may also be
non diagonal, and then Eq.(\ref{scalar2l}) does not hold in this
simple form.

In the case of the {\bf isotropic Heisenberg FP} $u_\perp^\star=u_\|^\star=u_\times^\star
=u^\star$ Eqs.(\ref{gapest1})-(\ref{gapast2}) simplify considerably. The ratios of the
elements of the $\mbox{\boldmath$\zeta$}_{\phi^2}$-matrix reduce to
\begin{equation}\label{isherat1}
\frac{\big[\mbox{\boldmath$\zeta$}_{\phi^2}^\star\big]_{12}}
{\zeta_+^\star-\big[\mbox{\boldmath$\zeta$}_{\phi^2}^\star\big]_{22}}=1
=\frac{\gamma_\|^\star}{\gamma_\perp^\star}
\end{equation}
for {\bf Case a} and
\begin{equation}\label{isherat2}
\frac{\big[\mbox{\boldmath$\zeta$}_{\phi^2}^\star\big]_{21}}
{\zeta_-^\star-\big[\mbox{\boldmath$\zeta$}_{\phi^2}^\star\big]_{11}}
=-\frac{n_\|}{n_\perp}=\frac{\gamma_\perp^\star}{\gamma_\|^\star}  \, ,
\end{equation}
for {\bf Case b}. For the second equalities (\ref{vgpagpe}) and (\ref{vgpegpa}) have been used.
Together with the relations $\varepsilon-2\zeta_+^\star=\alpha/\nu$ and
$\varepsilon-2\zeta_-^\star=\phi/\nu-d$, which introduce the critical exponents and
follow from Eqs. (80), (81) and (82)
in paper I, the values for the isotropic Heisenberg FP are:\\
{\bf Case a:} $\gamma_+^\star\ne 0$, $\gamma_-^\star=0$:
\begin{equation}\label{ihfix1}
\gamma_\perp^{\star 2}=\gamma_\|^{\star 2}=\frac{2}{n_\perp+n_\|}\frac{\alpha}{\nu} \, .
\end{equation}
{\bf Case b:} $\gamma_+^\star=0$, $\gamma_-^\star\ne 0$:
\begin{equation}\label{ihfix21}
\gamma_\perp^{\star 2}=\frac{2}{n_\perp+n_\|}\frac{n_\|}{n_\perp}\left(2\frac{\phi}{\nu}-d\right)  \, ,
\end{equation}
\begin{equation}\label{gapast22}
\gamma_\|^{\star 2}=\frac{2}{n_\perp+n_\|}\frac{n_\perp}{n_\|}\left(2\frac{\phi}{\nu}-d\right)  \, .
\end{equation}
Note that due to the sign in Eq.(\ref{isherat2}) the relation
\begin{equation}\label{gpagpepro}
\gamma_\|^\star\gamma_\perp^\star=-\frac{2}{n_\perp+n_\|}\left(2\frac{\phi}{\nu}-d\right)
\end{equation}
holds in this case. For $n_\|=1$ and $n_\perp=2$ our results agree with those of Ref.\cite{dohmmulti83}.

\begin{figure}[t,b]
      \centering{
       \epsfig{file=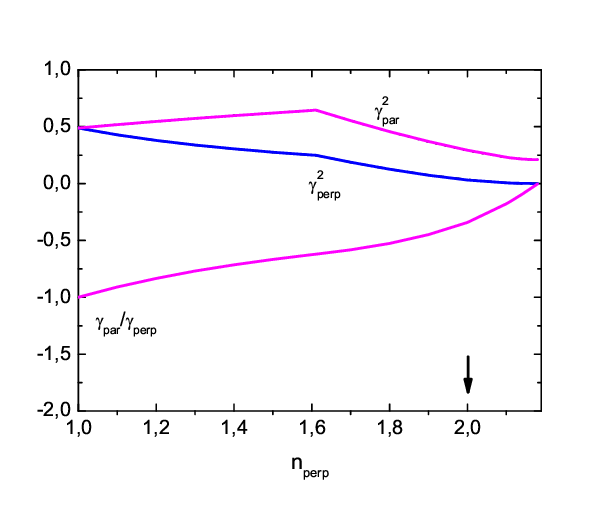,width=9cm,angle=0}  }
     \caption{ \label{gammacouplings} Fixed point values of the asymmetric static couplings
$\gamma_\|$ and $\gamma_\perp$  for {\bf Case b} ($\gamma_\perp^\star=0$, $\gamma_\|^\star\neq 0$)
at the Heisenberg FP  ($n_\perp<1.61$) and the biconical FP ($1.61<n_\perp<2.18$).}
   \end{figure}

\subsubsection{Resummation procedure}
As in our former papers\cite{partI,partII} of this series, in order to get numerical  estimates
we proceed within fixed dimension RG technique, i.e. we evaluate RG
expansions in couplings $\{u_\|,u_\perp,u_\times\}$ at  fixed $d=3$.
Furthermore, as far as the expansions are known to have zero radius
of convergence we use resummation technique\cite{resum} to get reliable
numerical estimates. The results given below were obtained within
such a technique applied to the two-loop RG expansions. One of the
ways to judge about typical numerical accuracy of our data, is to
give an estimate for some cases where the expansions (and,
subsequently, their numerical estimates) are  known within much
higher order of loops. As far as the static exponents $\alpha$ and
$\nu$ explicitly enter many of formulas considered above, let us
take them as an example. Namely, let us estimate relations
\begin{eqnarray}\label{l1}
(2\phi/\nu - d)|_{d=3} &\equiv& 2/\nu_- -3, \, ,\\ \label{l2}
\alpha/\nu|_{d=3} &\equiv& 2/\nu_+ -3 \, ,
\end{eqnarray}
that enter the formulas for the couplings $\gamma_\perp$,
$\gamma_\|$. The exponents $\nu_+$ and $\nu_-$ have been defined in Eqs. (80)
and (81) of paper I. Fig. \ref{exponents} shows  the dependence of $2\phi/\nu
- 3$ and of $\alpha/\nu$ on the order parameter component numbers
$n_\perp$ at fixed $n_\|=1$. Recall that of main interest for us
will be the physical case $n_\|=1$, $n_\perp=2$ indicated by the arrow. The region of
$n_\perp$ shown in the figure covers also the region of stability of
the Heisenberg $O(n)$-symmetrical FP, with $n=n_\|+n_\perp$. In
particular it starts at $n_\perp=1$ near the marginal field
dimension $n_c$ at which the exponent $\alpha$ changes its sign. For
the $O(n)$ vector model an estimate based on the fixed $d=3$ six
loop RG expansions reads \cite{Bervillier86}: $n_c=1.945\pm 0.002$.
We get for the correlation length critical exponent in the
Heisenberg $O(2)$ FP: $\nu=0.684$, via hyperscaling relation this
leads to $\alpha=-0.053$. Our estimate correctly reproduces the
absence of a divergency in the specific heat of $O(2)$ model
($\alpha$ is negative), however the value of $n_c\simeq 1.6$ we get
is rather underestimated. Note however, that the fixed $d$ approach
we exploit in two loop approximation is essentially better than the
corresponding $\varepsilon$ expansion. Indeed, in two-loop
$\varepsilon$-expansion one gets: $n_c=4-4\varepsilon$, which does
not lead to reasonable estimates \cite{note1}. The two loop estimate
of the massive field theory at $d=3$, $m_c\simeq 2.01$ \cite{Jug83},
is more close to the most accurate value of Ref.\cite{Bervillier86},
however it gives a wrong sign for the exponent $\alpha$. In any case
the negative value of $\alpha$ for $n_\perp=2$  agrees with other
calculations as reported in paper I.

It turns out (see below) that {\bf Case b} is the stable FP for the
asymmetric couplings. In order to evaluate numerically the values of
couplings $\gamma_\|^2$, $\gamma_\perp^2$, we therefore substitute
the resummed fixed point values of the static couplings $\{u_\|,u_\perp,u_\times\}$
into formulas (\ref{gapest2}), (\ref{gapast2})  and
resum the resulting expression. In principle, one can use different
ways for such an evaluation. Indeed, as we proceeded before, one can
present these formulas in the form of expansion in renormalized
couplings (keeping the two-loop terms) and resum the resulting
second-order polynomial. Alternatively, based on the observation
that numerator and denominator of Eqs. (\ref{gapest2}),
(\ref{gapast2}) contain combinations of critical exponents, one can
resum the numerator and the denominator separately. We will exploit
both ways which naturally will lead to slightly different numerical
estimates. This difference may also serve to get an idea about
typical numerical accuracy of the results. Separately, we will
evaluate the ratios $\gamma_\perp/\gamma_\|$. Again, it will be done
by resummation of the series for this ratio, Eq. (\ref{vgpegpa}), as
well as by using resummed values for $\gamma_\|^2$,
$\gamma_\perp^2$.In particular, for $n_\|=1$, $n_\perp=2$ we get:
$(\gamma_\perp^\star)^2=0.034$, $(\gamma_\|\star)^2=0.286$ (when
denominator and numerator are resummed separately), and
$(\gamma_\perp^\star)^2=0.031$, $(\gamma_\|\star)^2=0.293$ (when an
entire expression is resummed). Resulting differences of the order
of several percents bring about a typical numerical accuracy of the
estimates. In Fig. \ref{gammacouplings} we plot the FP values of the
asymmetric couplings and their ratio obtained  within
resummation of the entire expressions. These values will be used below to
calculate the critical dynamics.

\subsection{Static transient exponents}

The stability of the fixed points is determined by the sign of the corresponding transient
exponents. Latter can be found from the eigenvalues of the matrix
\begin{equation}\label{dbgadga}
\frac{\partial \beta_{\gamma_{\alpha}}}{\partial\gamma_{\beta}}
\end{equation}
with $\alpha,\beta=\perp,\|$. Inserting (\ref{betagamend}) into (\ref{dbgadga}) the
corresponding eigenvalues read
\begin{eqnarray}\label{lambdapm}
\lambda^{(\pm)}=\frac{1}{2}\Bigg\{-\varepsilon+n_\perp\gamma_\perp^2+n_\|\gamma_\|^2
+\big[\mbox{\boldmath$\zeta$}_{\phi^2}\big]_{11}+\big[\mbox{\boldmath$\zeta$}_{\phi^2}\big]_{22}
\nonumber \\
\pm\Bigg[\Bigg(\frac{n_\perp\gamma_\perp^2-n_\|\gamma_\|^2}{2}
+\big[\mbox{\boldmath$\zeta$}_{\phi^2}\big]_{11}-\big[\mbox{\boldmath$\zeta$}_{\phi^2}\big]_{22}\Bigg)^2
\nonumber \\
+\big(n_\perp\gamma_\perp\gamma_\|+2\big[\mbox{\boldmath$\zeta$}_{\phi^2}\big]_{12}\big)
\big(n_\|\gamma_\perp\gamma_\|+2\big[\mbox{\boldmath$\zeta$}_{\phi^2}\big]_{21}\big)\Bigg]^{1/2}\Bigg\} \, .
\end{eqnarray}
The above eigenvalues are valid in two loop order because Eq.(\ref{scalar2l}) already has been used.
The transient exponents
\begin{equation}\label{omgampm}
\omega_\gamma^{(\pm)}\equiv\lambda^{(\pm)}\Big(\{u\}=\{u^\star\},\{\gamma\}=\{\gamma^\star\}\Big)
\end{equation}
are calculated
by inserting the fixed point values of the static couplings into the eigenvalues.
With the fixed point values (\ref{ihfix1}) - (\ref{gapast22}) and Eq.
(\ref{lambdapm}) we obtain for the isotropic Heisenberg FP the transient exponents \\
{\bf Case a:} $\gamma_+^\star\ne 0$, $\gamma_-^\star=0$:
\begin{equation}\label{transplus}
\omega^{(+)}=\frac{\alpha}{\nu}  \ , \qquad  \omega^{(-)}= -\bar{W}^\star  \, .
\end{equation}
{\bf Case b:} $\gamma_+^\star=0$, $\gamma_-^\star\ne 0$:
\begin{equation}\label{transminus}
\omega^{(+)}=2\frac{\phi}{\nu}-d  \ , \qquad  \omega^{(-)}= \bar{W}^\star   \, .
\end{equation}
$\bar{W}^\star$ is the root
\begin{equation}\label{barw}
\bar{W}\equiv\sqrt{\Big(\big[\mbox{\boldmath$\zeta$}_{\phi^2}\big]_{11}
-\big[\mbox{\boldmath$\zeta$}_{\phi^2}\big]_{22}\Big)^2
+4\big[\mbox{\boldmath$\zeta$}_{\phi^2}\big]_{12}\big[\mbox{\boldmath$\zeta$}_{\phi^2}\big]_{21}}
\end{equation}
taken at the fixed point values of the couplings. It is always positive and
reads at the isotropic Heisenberg FP in two loop order
\begin{equation}\label{barw2l}
\bar{W}^\star=\frac{n_\perp+n_\|}{6}u^\star\left(1-\frac{u^\star}{3}\right)    \, .
\end{equation}
Thus one concludes that {\bf Case b} is the stable FP even if
$\alpha$ would be positive\cite{rem2}. For the biconical FP the
stability of {\bf Case b} can be verified explicitly by the flow of
the couplings.

\subsection{Dynamical fixed points}

Calculations of the dynamical FPs values are done by solving the
FP equations for the dynamical $\beta$-functions. Since only two of the equations
(\ref{betav})-(\ref{betawperp}) are independent
the third equation serves as consistency check of the solution found.
It is useful to choose for this purpose Eqs (\ref{betawpara}) and (\ref{betawperp}) for the time
scale ratios $w_\|$ and $w_\perp$. Then one has to solve
\begin{eqnarray}  \label{dynFPs}
 \beta_{w_\|}(w_\|,w_\perp,w_\|/w_\perp) &=& 0 \,,  \\
 \beta_{w_\perp}(w_\|,w_\perp,w_\|/w_\perp) &=& 0 \,  .  \nonumber
\end{eqnarray}
To find the dynamical FP values coordinates,  the resummed FP values of static couplings
$u_\|^*,u_\perp^*,u_\times^*,\gamma_\perp^*,\gamma_\|^*$ are inserted into these
equations\cite{note2}.

The dynamical FPs depend on which static FP is considered. There might be several dynamical FPs for one static
FP, which could be  either strong dynamical or weak dynamical scaling FPs. Since also unstable static
FPs might be reached in the asymptotics if one  starts with static initial conditions in the attraction
region of this FP (a subspace in the space the static couplings see e.g.
Fig. 3 in paper I) at least both the Heisenberg FP and the biconical FP have to be taken into consideration.
It turns out that in both cases apart from the trivial unstable FP where all timescale ratios are zero
(see below) two dynamical FPs are found: (i) an unstable weak dynamical scaling FP corresponding to model A
and (ii) a stable  new strong dynamical scaling FP. In the physical interesting case $n_\|=1$ and
$n_\perp=2$ these cases correspond to dynamical behavior at a multicritical point of bicritical and
tetracritical type respectively. The different types of weak and strong dynamical FPs are shown in Tab. \ref{tab2}.

\begin{table*}[t]
\centering \tabcolsep=3mm
\begin{tabular}{llllllll}
 \hline \hline
   FP & scaling type & $v$  & $w_\|$ & $w_\perp= w_\| /v$ & $z_{\phi_\|}$ & $z_{\phi_\perp}$ & $z_{m}$ \\ \hline \hline \\
$\mathcal{H}_{C_w}$ & weak &  0  & $w_{\| {C_w}}^{\mathcal{H}}$   & $\infty$ & $2\frac{\phi}{\nu}-1$ &
infinite fast
&$2\frac{\phi}{\nu}-1$
\\
$\mathcal{H}_A$ & weak &  $v_A^{({\mathcal H})}$  & 0  & 0 & $2+c\eta$ & $2+c\eta$ & $2\frac{\phi}{\nu}-1$\\
$\mathcal{H}_C$ & strong & $v_C^{\mathcal{H}}$  & $w^{\mathcal{H}}_\|$
&  $w^{\mathcal{H}}_\perp$  & $2\frac{\phi}{\nu}-1$ & $2\frac{\phi}{\nu}-1$ & $2\frac{\phi}{\nu}-1$ \\ \\ \hline \\
 $\mathcal{B}_{C_w}$& weak   & 0 &  $w_{\| {C_w}}^{\mathcal{B}}$ & $\infty$  & $2\frac{\phi}{\nu}-1$  & infinite fast
&$2\frac{\phi}{\nu}-1$  \\
 $\mathcal{B}_A$   & weak   &  $v_{A}^{({\mathcal B})}$ &  $0$ & $0$  & $z^{\mathcal{B}}$  & $z^{\mathcal{B}}$ &
$2\frac{\phi}{\nu}-1$ \\
$\mathcal{B}_C$ & strong &  $v_C^{({\mathcal B})}$ & $w^{\mathcal{B}}_\|$ & $w^{\mathcal{B}}_\perp$  & $2\frac{\phi}{\nu}-1$
& $2\frac{\phi}{\nu}-1$  & $2\frac{\phi}{\nu}-1$  \\ \\
 \hline \hline
\end{tabular}
\caption{Types of dynamical FPs for the static Heisenberg and biconical FP. Not included is the trivial unstable fixed point
with  all times scale ratios equal to zero. The value of $c$ reads $c=6\ln (4/3)-1$. Note that for the weak scaling FP the
result
is only valid in two loop order, whereas the relation of the dynamical critical exponent in the strong scaling FP holds in all
orders. \label{tab2}}
\end{table*}

\subsubsection{Strong dynamical scaling fixed point}

\begin{table*}[h,t,b]
\centering \tabcolsep=3mm
\begin{tabular}{lllllllll}
 \hline \hline
 FP & $u_\|^\star$ & $u_\perp^\star$ & $u_\times^\star$ & $\gamma_\|^{\star 2}$ & $\gamma_\perp^{\star 2}$
& $v^\star$ & $w_\perp^\star$    & $w_\|^\star$ \\ \hline
   ${\mathcal B_C}$, {\bf a}  & 1.28745 & 1.12769 & 0.30129 & 0.29378 & 0.03170
 & $6.09592\cdot10^{-43}$ & $1.24285\cdot10^{42}$  & 0.75763    \\
    ${\mathcal H_C}$, {\bf b} &  &  &  &
 &  & 7.29393$\cdot 10^{-5}$ & 1.55665$\cdot10^{4}$  & 1.13541       \\
   ${\mathcal H_C}$,  {\bf a} & \raisebox{1.5ex}[-1.5ex]{$1.00156$}  &
\raisebox{1.5ex}[-1.5ex]{$1.00156$}
& \raisebox{1.5ex}[-1.5ex]{$1.00156$}  & \raisebox{1.5ex}[-1.5ex]{$0.72554 $}
 & \raisebox{1.5ex}[-1.5ex]{$0.18139 $}   & 7.30771$\cdot 10^{-5}$ & 1.55372$\cdot10^{4}$  & 1.13541 \\
     \hline \hline
\end{tabular}
\caption{\label{tab3} FP values of  couplings and timescale ratios for $n_\|=1$,
$n_\perp=2$. {\bf a} FP values of the timescale ratios found
via approximation using Eqs. (100), (101), as described in the text with the values for
FP ${\mathcal B}$: $A=0.09770$, $B=0.00101$ and the FP ${\mathcal H}$:
$A=0.31534$, $B=0.03311$. {\bf b} numerical solution for the FP values of the timescale ratios;  }
\end{table*}
In order to find the strong dynamical scaling FPs it is not necessary to discriminate between the
static Heisenberg FP or the biconical FP although the dynamical equations to be solved simplify in the
first case a little bit. Thus we use the results for the FP values derived in paper I  for the quartic
couplings $\{u\}$ and the FP values for the asymmetric couplings $\{\gamma\}$ of {\bf Case b} (see Eqs.
(\ref{vgpegpa}), (\ref{gapest2}) and (\ref{gapast2})). At the strong scaling dynamical FP all timescale
ratios have to be nonzero and finite. Moreover due to the definitions
 of the timescale  ratios it follows that  $w_\|^\star=v^\star w_\perp^\star$. This dynamical strong
scaling FP value is found by setting  the differences  of two of the three dynamical $\zeta$-functions,
(\ref{zetalambda})-(\ref{zetagapara2l}) to zero leading to three equations. Since  there are only
two independent time ratios the third equation can be used to check the results.

\begin{figure}[t,b]
      \centering{
       \epsfig{file=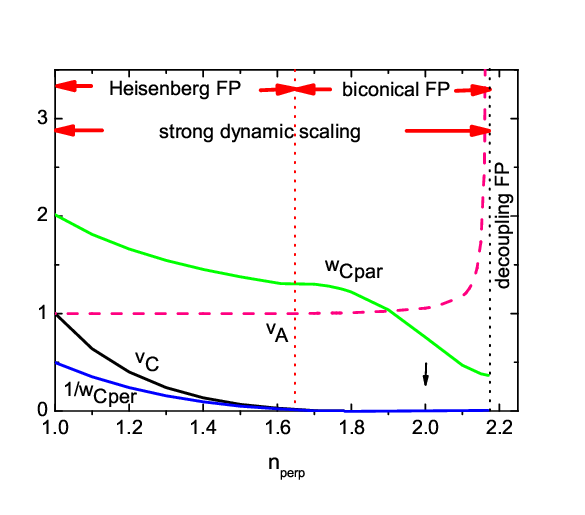,width=9cm,angle=0}  } \vspace{-1.0cm}
     \caption{ \label{HDFPP} Fixed point values of the timescale ratios
$v$, $1/w_\perp$ and $w_\|$  for the static stable FPs: the
isotropic Heisenberg FP ($n_\perp<1.61$) and the biconical FP
($1.61<n_\perp<2.18$). Strong dynamical scaling is valid up to the
stability borderline to the decoupling FP. In the biconical region
the values of $v_C$ and $1/w_{C\perp}$ are finite but not to
distinguish from zero on this scale. The notation of the dynamical FPs
correspond to the notation in Tab \ref{tab2}. The dashed curve shows
the unstable model A FP (see text).}
\end{figure}

The FP values (FP with subscript $C$ in Tab. \ref{tab2}) have been plotted in Fig. \ref{HDFPP} for different $n_\perp$ at
$n_\|=1$ and the numerical values for $n_\perp=2$ are collected in Tab. \ref{tab3}. This shows that the FP value of the
timescale ratio for  $v$ is different from the FP value found in the pure relaxational model A. These were at the Heisenberg FP
$v^\star_{A}=1$ and at the biconical FP $v^\star=v^{\mathcal{B}}_{A}$ with $v^{\mathcal{B}}_{A}\to\infty$ in approaching the
stability borderline to the decoupling fixed point (see Fig. 1 in paper II).

\begin{figure} [h,t,b]
      \centering{
       \epsfig{file=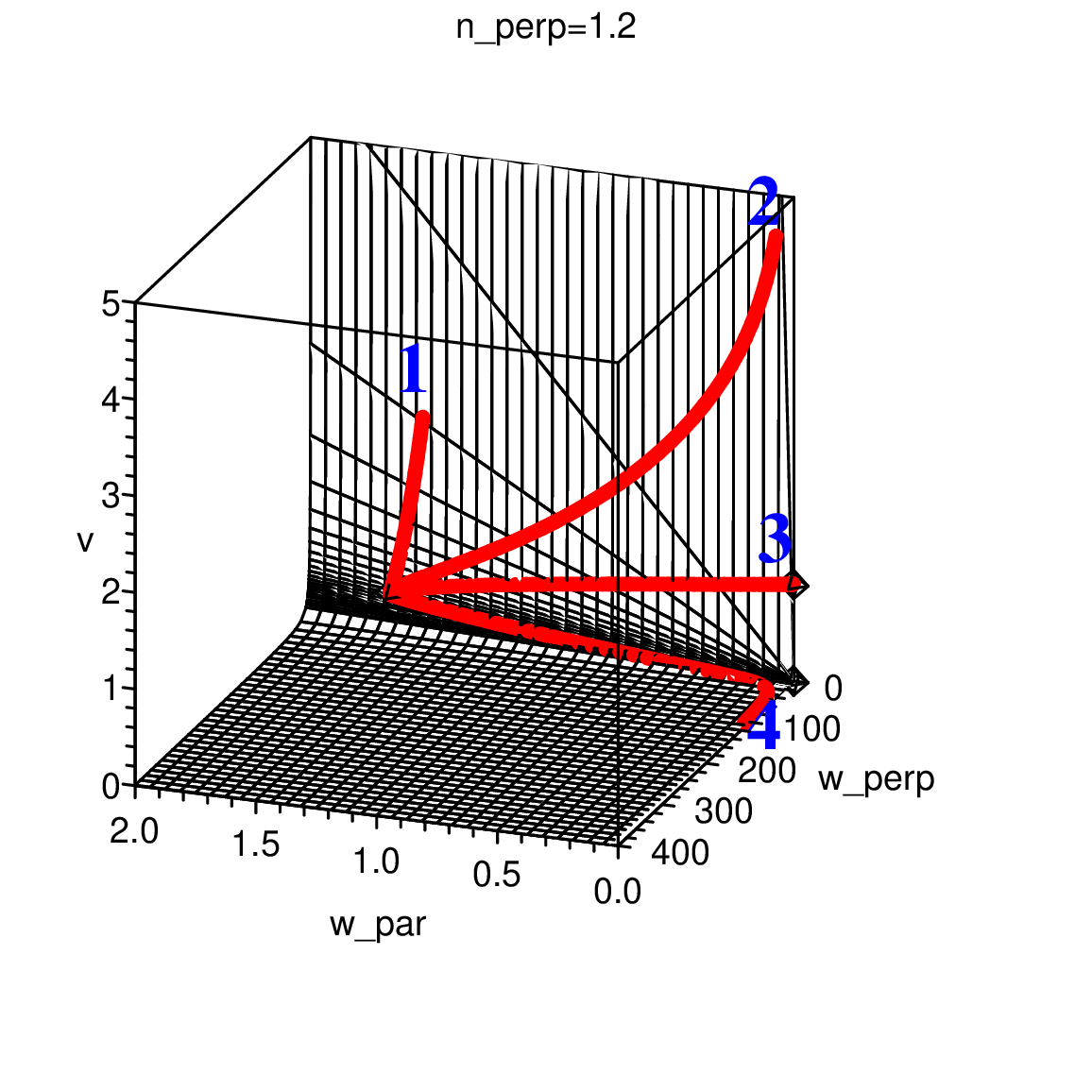,width=9cm,angle=0}
       \epsfig{file=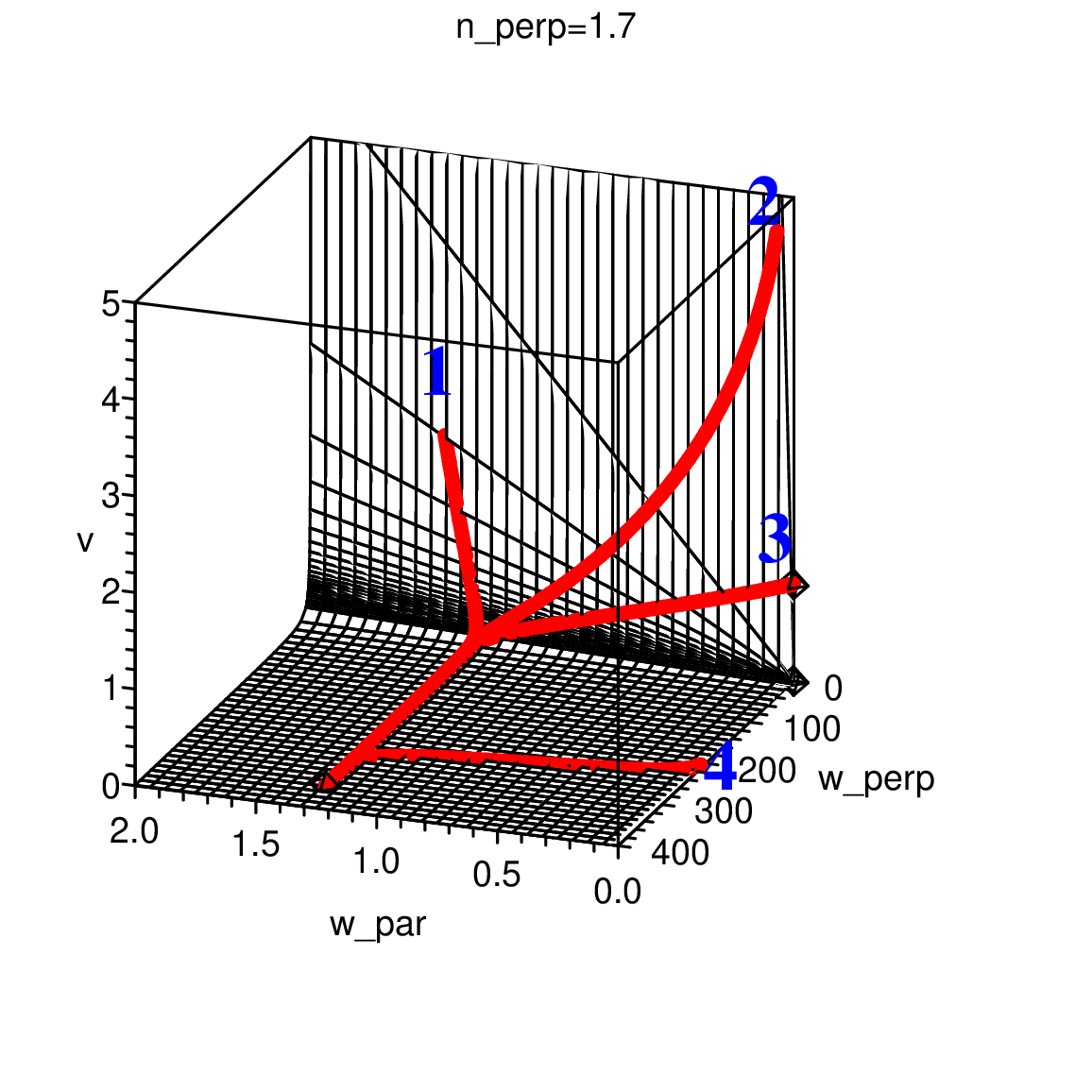,width=9cm,angle=0}
        }  \vspace{-1.57cm}
     \caption{ \label{flow12} Dynamical flow at $n_\|=1$ and different $n_\perp$ values for different dynamical initial 
conditions numbered 1 to 4. The static couplings are chosen
to be fixed at their stable FP values (the isotropic Heisenberg FP for $n_\perp=1.2$, the biconical FP for $n_\perp=1.7$).
The dynamical FP values are $v^\star=0.399$, $0.004$,  $w^\star_\|=1.661$, $1.300$ and $w^\star_\perp=4.159$,
$351.06$ at ${\cal H_C}$, ${\cal B_C}$ respectively. Also shown is the surface $v=w_\|/w_\perp$ to which the flow is
restricted.}
   \end{figure}

   \begin{figure}[h,t,b]
      \centering{
       \epsfig{file=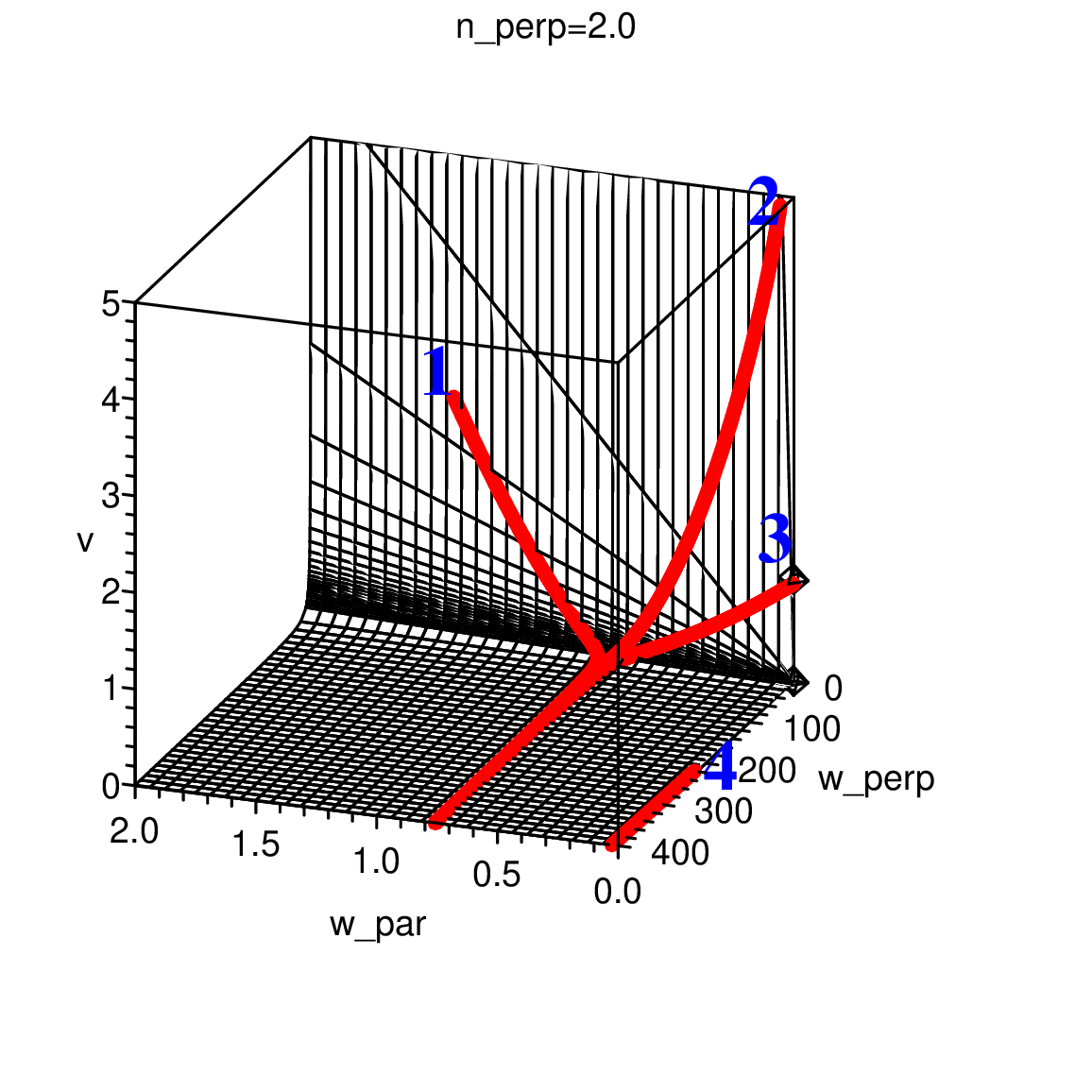,width=9cm,angle=0}  }  \vspace{-1.5cm}
     \caption{ \label{flow02} Dynamical flow at $n_\|=1$ and $n_\perp=2$ for different dynamical initial 
conditions numbered 1 to 4. The static couplings are chosen to be fixed at 
their  biconical
FP values. The static and dynamical FP values of ${\cal B_C}$ are given in Tab. \ref{tab3}. The dynamical FP lies outside the
region shown. Also shown is the surface $v=w_\|/w_\perp$ to which the flow is
restricted.}
   \end{figure}

A numerical problem arises in finding the FP values of $v$ and
$1/w_\perp$when they reach very small values. It cannot be
numerically decided wether the FP values are zero or finite. In
order to clarify the existence or nonexistence of a weak scaling FP
one has to look for an analytic expression for the small FP values.
However it is numerically easy to find the FP value of $w_\|$ which
is nonzero and finite in the whole region up to the stability
borderline between the biconical and decoupling FP. In order to
solve this problem the dependence of the $\zeta$-functions are
studied within this region. One observes  that there are logarithmic
terms which would diverge in the limit $v\to 0$ under the condition
$w_\perp=w_\|/v$. Thus one obtains two equations for the FP value of
$v$ and $w_\|$. In the equation for the FP of $w_\|$ one might
safely perform the limit $v\to 0$ and $w_\perp \to \infty$. This
leads to
\begin{equation}\label{dynfppar}
0=\zeta_{\Gamma_\|}^{(C)}\big(\{u^\star\},\{\gamma^\star\},v=0, w_\|, w_\perp\to\infty\big)-2\zeta^\star_m  \, .
\end{equation}
Using limiting functions
\begin{equation}
\zeta_{\Gamma_\|}^{(A)}=\frac{n_\|+2}{36}\ u_\|^{\star 2}\left(3\ln\frac{4}{3}
-\frac{1}{2}\right)  -\frac{n_\perp}{72}\ u_\times^{\star 2}
\end{equation}
and
\begin{eqnarray}
\zeta_{\Gamma_{\|}}^{(C_{\|})}\big(u^\star_{\|},\gamma^\star_{\|},
w_{|}\big)=\frac{w_{\|}\gamma_{\|}^{\star 2}}{1+w_{\|}}\Bigg\{1
-\frac{1}{2}\Bigg[\frac{n_{\|} +2}{3}\nonumber \\
u^\star_{\|}\Big(1-3\ln\frac{4}{3}\Big)
+\frac{w_{\|}\gamma_{\|}^{\star 2}}{1+w_{\|}}\Bigg(\frac{n_{\|}}{2}
-\frac{w_{\|}}{1+w_{|}}-  \\
\frac{3(n_{\|}+2)}{2}\ln\frac{4}{3}
-\frac{(1+2w_{\|})}{1+w_{\|}}\ln\frac{(1+w_{\|})^2}{1+2w_{\|}}
\Bigg)\Bigg]\Bigg\}  \nonumber
\end{eqnarray}
then Eq. (\ref{dynfppar}) reads
\begin{eqnarray}
\label{dynfpsmall}
0=\frac{w_{\|}\gamma_{\|}^{\star 2}}{1+w_{\|}}\Bigg\{1
-\frac{1}{2}\Bigg[\frac{n_{\|}
+2}{3}u^\star_{\|}\Big(1-3\ln\frac{4}{3}\Big)\nonumber \\
+\frac{w_{\|}\gamma_{\|}^{\star 2}}{1+w_{\|}}\Bigg(\frac{n_{\|}}{2}
-\frac{w_{\|}}{1+w_{\|}}-\frac{3(n_{\|}+2)}{2}\ln\frac{4}{3}
\nonumber \\
-\frac{1+2w_{\|}}{1+w_{\|}}\ln\frac{(1+w_{\|})^2}{1+2w_{\|}}
\Bigg)\Bigg]\Bigg\} \\
-\frac{n_\perp}{4}\frac{w_\|\gamma^\star_\|\gamma^\star_\perp}{1+w_\|}\Bigg[
\frac{2}{3}u_\times+\frac{w_\|\gamma^\star_\|\gamma^\star_\perp}{1+w_\|}\Bigg]
\nonumber \\
+\frac{n_\|+2}{36}\ u_\|^{\star 2}\left(3\ln\frac{4}{3}
-\frac{1}{2}\right)  -\frac{n_\perp}{72}\ u_\times^{\star 2}-2\zeta^\star_m  \nonumber
\end{eqnarray}
This equation is solved numerically to give the value of $w_\|^\star$ which then is inserted into
the second equation for $v^\star$. In order to find  $v^\star$ one collects the logarithmic diverging terms
in the equation for $v^\star$,
\begin{equation}
\zeta_{\Gamma_\|}^{(C)}(v,w_\|,w_\|/v)-\zeta_{\Gamma_\perp}^{(C)}(v,w_\|,w_\|/v)=0 \, .
\end{equation}
In the remaining terms the limit $v\to 0$ can safely performed.
Then the solution reads
\begin{equation}
\ln v^\star=-\frac{A}{B}
\end{equation}
with
\begin{eqnarray}  \label{Aslow}
A=2\zeta^\star_m -\gamma_\perp^{\star 2}\Bigg\{1
-\frac{1}{2}\Bigg[\frac{n_{\perp}+2}{3}u^\star_{\perp}\Big(1-3\ln\frac{4}{3}\Big)\nonumber \\
+\gamma_{\perp}^{\star 2}\Bigg(\frac{n_{\perp}}{2}
-1-\frac{3(n_{\perp}+2)}{2}\ln\frac{4}{3}-2\ln\frac{w^\star_\|}{2}\Bigg)\Bigg]\Bigg\} \nonumber \\
+\frac{n_\|}{4}\gamma^\star_\perp\gamma^\star_\|
\Bigg[\frac{2}{3}u^\star_\times+\gamma^\star_\perp\gamma^\star_\|\Bigg]\ln2
-\frac{n_\perp+2}{36}\ u_\perp^{\star 2} \nonumber \\
\left(3\ln\frac{4}{3}
-\frac{1}{2}\right)
-\frac{n_\|}{72}\ u_\times^{\star 2}\Big(1-2\ln 2\Big)
\end{eqnarray}
and
\begin{equation}\label{slow}
B=\gamma_\perp^{\star 4}+\frac{n_\|}{36}u_\times^{\star
2}+\frac{n_\|}{4}\gamma^\star_\perp\gamma^\star_\|\Bigg[\frac{2}{3}u^\star_\times+\gamma^\star_\perp\gamma^\star_\|\Bigg] \, 
.
\end{equation}
It can be shown that $A$ and $B$ are positive. Approaching the stability borderline to the decoupling FP $A$ stays finite and $B$ goes to zero
 since $u^\star_\times$ and $\gamma^\star_\perp$ go to zero. In consequence $v^\star$ goes to zero and $w^\star_\perp$ goes to infinity in
and only in this limit.  The analytic solution found within this region joins smoothly  to the numerical solution found for larger
values of the timescale ratios. Thus it is proven that in the whole region where the Heisenberg FP or the biconical FP is stable
dynamical strong scaling holds.

Considering the FP values for the timescale ratios for the Heisenberg FP in the region of $n_\perp>1.7$ (where it is
reached only for static initial conditions in a subspace of the fourth order couplings) one also finds a small value
for $v^\star$, a very large value for $w_\perp^\star$ and a nonzero finite value for $w_\|^\star$. However contrary to the
biconical FP now $A$ and $B$ stay finite at the stability borderline between the biconical FP and the decoupling FP at
$n_\perp\sim2.18$. Indeed the values calculated for the Heisenberg FP from Eqs (\ref{Aslow}) and (\ref{slow}) are
$A=0.34866$, $B=0.02558$ whereas for the biconical FP one obtains $A=0.05268$, $B=4.27958\cdot 10^{-9}$.

The asymptotic dynamical exponents are obtained from the values of the $\zeta$-functions at the FP:
\begin{equation}
z_{\phi_\|}=2+\zeta^\star_{\Gamma_\|}  \qquad   z_{\phi_\perp}=2+\zeta^\star_{\Gamma_\perp}
\qquad z^\star_{m}=2+\zeta_{\lambda} \, .
\end{equation}
At the strong dynamical scaling FP all dynamical $\zeta$-functions are equal to twice the static
$\zeta$-function $\zeta_m$. Therefore the CD induces the value of the dynamical critical exponent $z$
\begin{equation} \label{zstrong}
z=2+2\zeta_m^\star=2\frac{\phi}{\nu}-1
\end{equation}
according to Eq. (\ref{zetam}) and Eq. (\ref{gmifix}). The values
for the static exponents depend on which static FP is stable. For
$n_\|=1$ the $n_\perp$-dependence of $z-2$ is shown in Fig.
\ref{exponents}.

\subsubsection{Weak dynamical scaling fixed point}

Weak dynamical scaling FPs are solutions of the dynamical FP equations  where one or more of the FP values of
the timescale ratios are zero or infinite. Such a weak dynamical scaling FP has already been found in model A
and it became stable  at the stability borderline to the decoupling FP.

Indeed Eqs (\ref{dynFPs}) allow solutions where both
timescale ratios $w_\|$ and $w_\perp$ are zero. In such a case one
has to rely on the third equation for the ratio $v=w_\|/w_\perp$ to
find the limiting FP value. However in the limit $w_\|\to 0$ and
$w_\perp\to 0$ Eq. (\ref{betav}) for $v$ reduces to the FP equation
of model A (FP with subscript $A$ in Tab. \ref{tab2}). Thus one
recovers the model A FPs in this case.

There is no solution $w^\star_\|=v^\star=0$ and $w_\perp$ nonzero
and finite  due to the $\ln v$ term in (\ref{zetagperp}).  For a
similar reason no FP with $w^\star_\|$ nonzero and finite,
$w^\star_\perp=0$ and $v^\star=\infty$ is possible. However a FP
with  $w^\star_\|$ nonzero and finite, $v^\star=0$ and
$w_\perp=\infty$ is possible (FP with subscript $C_w$ in Tab.
\ref{tab2}). The values of $w_\|^\star$ are obtained from Eq.
(\ref{dynfpsmall}), but now these values are not an approximation
but the exact $C_w$ FP values for any $n_\|$ and $n_\perp$.

The dynamical critical exponents may be different in the case of weak dynamical scaling. For the weak model C FP
(subscript ${C_w}$) $w_\|^\star$ is finite and nonzero therefore $\zeta_{\Gamma_\|}^\star=\zeta_\lambda^\star$ and
$z_{\phi_\|}=2\frac{\phi}{\nu}-1$.  Thus the CD sets the timescale for the OP $\phi_\|$.
Inserting $w_\perp^\star=\infty$ into $\zeta_{\Gamma_\perp}$ leads due to logarithmic
diverging terms to an infinite value of the corresponding dynamical exponent $z_{\phi_\perp}$. This indicates
that the density $\phi_\perp$ is much faster than the other densities. It is especially much faster than
the other OP $\phi_\|$.

In the case where both FP values of the timescale ratios $w_\|$ and $w_\perp$ are zero and $v$ 
is finite and nonzero, both OPs have the same timescale with the dynamical exponent (of model A) $z=2+c\eta$
different from the exponent of the CD $z_m=2\frac{\phi}{\nu}-1$.

\subsection{Dynamical transient exponents}

The dynamical transient exponents can be calculated from the matrix of
the derivatives of the $\beta$-function with respect to the
timescale ratios $v$, $w_\|$ and $w_\perp$. Since only two timescale
ratios are independent only two are considered in the stability
matrix. The eigenvalues of the $2\times 2$-matrix have to be
positive for the overall stable FP otherwise the FP is unstable.  In
the following the timescale ratios $w_\|$ and $w_\perp$ are chosen
as independent.

The model A type FP with $v^\star=v_A$ nonzero and finite is unstable since the two eigenvalues,
\begin{equation} \label{omegas}
\omega_{w_\|}=\zeta^\star_{\Gamma_\|}-2\zeta^\star_m \qquad \mbox{and} \qquad
\omega_{w_\perp}=\zeta^\star_{\Gamma_\perp}-2\zeta^\star_m
\end{equation}
are negative. In fact they are equal because $\zeta^\star_{\Gamma_\|}=\zeta^\star_{\Gamma_\perp}$.
Their values are calculated with $2\zeta^\star_m=2\phi/\nu-1$ and inserting the model A FP value
$v^\star=v_A$ (see Fig \ref{HDFPP}) and $w_\|^\star=w_\perp^\star=0$. Then
$\zeta^\star_{\Gamma_\|}-2\phi/\nu+1<0$ and $\omega_{w_\|}=-0.126$ for  $n_\|=1$ and $n_\|=2$.

Similarly the instability of the FP with
$v^\star=w_\|^\star=w_\perp^\star=0$ can be shown. However some care
has to be taken  due to the vanishing timescale ratio $v$. The
eigenvalues are again given by Eq. (\ref{omegas}) but now they are
different. Whereas $\omega_{w_\|}$ is negative $\omega_{w_\perp}$
goes to $\infty$ due to the $\ln v$ term in the model A $\zeta$
function (\ref{zetagperp}).

The transient exponent for the strong scaling FP are the eigenvalues of the matrix of derivatives of the $\beta$-functions according to the timescale ratios at the FP
\begin{equation}
\left(\begin{array}{cc} \frac{\partial \beta_{w_\|}}{\partial w_\|}&  \frac{\partial \beta_{w_\|}}{\partial w_\perp}  \\
\frac{\partial \beta_{w_\perp}}{\partial w_\|}& \frac{\partial \beta_{w_\perp}}{\partial w_\perp}
\end{array}\right)^\star=
\left(\begin{array}{cc} w_\| \left(\frac{\partial \zeta_{\Gamma_\|}}{\partial w_\|}\right) & w_\| \left(\frac{\partial \zeta_{\Gamma_\|}}{\partial w_\perp}\right) \\
w_\perp\left( \frac{\partial \zeta_{\Gamma_\perp}}{\partial w_\|}\right) & w_\perp \left(\frac{\partial \zeta_{\Gamma_\perp}}{\partial w_\perp}\right)
\end{array}\right)^\star \, .\\
\end{equation}
Use has been made from the independence of $\zeta_\lambda$ on the
timescale ratios. The nondiagonal elements depend on the timescale
ratios by which they are derived only via $v$ and therefore are
proportional to $1/w_\perp$. In the region where $w_\perp$ is very
large the two eigenvalues are then given by the diagonal elements.
\begin{equation}
\omega_{w\|}=w_\|^\star \left(\frac{\partial \zeta_{\Gamma_\|}}{\partial w_\|}\right)^\star \, , \quad
\omega_{w_\perp}=w_\perp^\star \left(\frac{\partial \zeta_{\Gamma_\perp}}{\partial w_\perp}\right)^\star \, .
\end{equation}
Near the stability borderline to the decoupling FP the second eigenvalue goes to zero according to
\begin{equation}
\omega_{w_\perp}=B+{\cal O}(1/w_\perp)
\end{equation}
with $B$ from Eq. (\ref{slow}), being exactly zero at the
borderline. The value of the slow transient at $n_\|=1$ and
$n_\perp=2$ is given by $\omega_{w_\perp}=0.001$  Thus as shown in
the next section in this case nonasymptotic effects are present in
the physical accessible region.

As already mentioned for the Heisenberg FP  ${\cal H_C}$  $B$ does not reach zero at the borderline but its value is one
order smaller than the static transient exponents. At $n_\perp=2$ and $n_\perp=2.18$  the value of the dynamical transient
exponent is given by $\omega_{w_\perp}=0.033$ and $\omega_{w_\perp}=0.026$ respectively.

\section{Dynamical flows and effective exponents \label{flows}}

The flow of the timescale ratios is described by the RG equations
\begin{eqnarray}
l\frac{\partial v}{\partial l}&=& \beta_v (\{u\},\{\gamma\},\{w\})  \, , \nonumber \\
l\frac{\partial w_\perp}{\partial l}&=& \beta_{w_\perp} (\{u\},\{\gamma\}, \{w\}) \, ,  \\
l\frac{\partial w_\|}{\partial l}&=& \beta_{w_\|} (\{u\},\{\gamma\},\{w\})\, ,   \nonumber
\end{eqnarray}
with the $\beta$-functions Eqs.
(\ref{betav})-(\ref{betawpara}). Note, that the dynamical $\beta$-functions depend also on the RG equations of
the static quartic couplings (Eqs (33)-(36) of paper I) and the RG equations of the asymmetric couplings
\begin{equation}
l\frac{\partial \vec{\gamma}}{\partial l}=\vec{\beta}_\gamma(\{u\},\{\gamma\})
\end{equation}
with the $\beta$-function Eq. (\ref{betagamend}).

In order to simplify the picture it is assumed that the static couplings have already reached
the FP by which they are attracted from their initial conditions. Then
the  RG flows are displayed in the
three-dimensional space of the time-scale ratios $w_\|, w_\perp, v$
in Figs. \ref{flow12} and \ref{flow02} for the physical interesting case
at $n_\|=1$, $n_\perp=2$. The static biconical FP is
stable for this case in general. However for initial conditions on the surface separating
the stable biconical FP from the Heisenberg FP the flow is attracted to the Heisenberg FP.
Therefore one may also fix the static parameters to this FP.

To give an overview of the different patterns of the flows three
different value of $n_\perp$ are chosen for fixed $n_\|=1$: (i)
$n_\perp=1.2$, where the static Heisenberg FP is stable (Fig.
\ref{flow12}), (ii) $n_\perp=1.7$  (Fig. \ref{flow12}) and (iii)
$n_\perp=2$ (Fig. \ref{flow02}) where the biconical FP is  stable.

In all cases the FP values of the timescales are nonzero and finite but the values of $v^\star$
becomes very small and $w_\perp^\star$ very large. The asymptotic approach to the FP in cases (ii) and (iii)
occurs in the direction of the $w_\perp$-axis almost at $v\sim 0$ and $w_|\sim w^\star_\|$.

\subsection{Effective exponents \label{effex}}

\begin{figure}[h,t,b]
      \centering{
       \epsfig{file=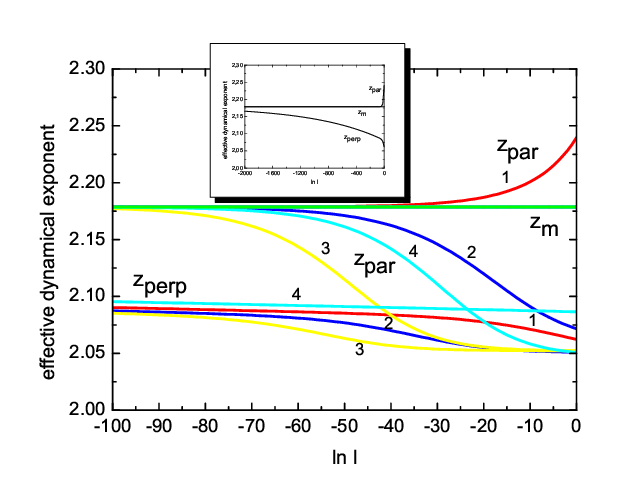,width=10cm,angle=0}}
\caption{Effective dynamical exponents $z_\|$,
$z_\perp$, and $z_m$  calculated along the
different RG flows of Fig. \ref{flow02} (indicated by the numbers). The insert shows that even for flow parameters
as small as $\ln l=-2000$ the effective exponent $z_\perp$ has not reached its asymptotic value $2.18$.
 \label{fig_eff_exp1}}
\end{figure}
We define the effective exponents by:
\begin{eqnarray} \nonumber
 z_{\|\, eff}(l)    &=& 2 + \zeta_{\Gamma_\|} (\{u(l)\},\{\gamma(l)\},\{w(l)\}), \\ \nonumber
 z_{\perp\, eff}(l) &=& 2 + \zeta_{\Gamma_\perp} (\{u(l)\},\{\gamma(l)\},\{w(l)\}), \\ \label{effexps}
 z_{m \, eff} (l)    &=& 2 + \zeta_\lambda (\{u(l)\}) \, .
\end{eqnarray}
These exponents appear e.g. in the critical temperature  and/or wave
vector dependence of the transport coefficients describing the
relaxation of the alternating magnetization or the diffusion of the
magnetization in the direction of the external magnetic field. They
are in principle experimentally accessible. An interesting feature
is the independence of the effective dynamical scaling exponent
$z_m$ of the CD from the dynamical timescales. Therefore its
nonasymptotic value is only due to nonasymptotic
effects within statics. This allows to trace back nonasymptotic
effects in dynamical quantities to the slow transients in statics or
those appearing in dynamics.

In order to calculate the numerical values of the effective exponents, we substitute into Eq.
(\ref{effexps}) the resummed coordinates of the static FP $\{u_\|^*, u_\perp^*, u_\times^*\}$ and
the values of timescale ratios $\{w_\|(\ell), w_\perp(\ell), v(\ell) \}$ along the RG
flow. Choosing the FP values of the static couplings fixes the asymptotic values of the effective dynamical exponents
since they are expressed by the static asymptotic exponents for the strong dynamical scaling FP.

The results shown in Fig. \ref{fig_eff_exp1} correspond to the case $n_\|=1$,
$n_\perp=2$. We evaluate the timescale ratios along the previously
obtained flows 1-4 in Fig. \ref{flow02}. The effective exponents for the different initial conditions
are shown by numbered solid lines,
As one can observe from this figure, the exponents calculated along several flows  do not coincide
for the values of the flow parameter shown. However one sees the merging of the different values for $z_\|$ to their
asymptotic value $z_\perp^\star=z_m^\star=2.18$ given by the static value corresponding to the CD (the constant line in Fig.
\ref{fig_eff_exp1}). More remarkable is the difference between the effective exponents for the parallel and perpendicular components of
the OP. This difference can be traced back to the fact that $w_\perp$ did not reach its very large FP value due to the slow
dynamical transient. Only for flow parameter values larger than $\ln l\sim -2000$ the effective exponent attains its
asymptotic
value $z_\perp\sim 2.18$ as it should be (see the insert of Fig. \ref{fig_eff_exp1} where the smallest value is $\ln l=
-2000$ for the flow 1).

Furthermore, we see that the exponent $z_\|$ flows towards its
asymptotic value $z_\| =2.18$. In fact, exponent $z_\perp$ attains
the asymptotic  value $z_\perp=2.18$ as well, but for much smaller
values of the flow parameter. In the insert of Fig. \ref{fig_eff_exp1} we show
this exponent for all flows within larger range of $\ln \ell$.

For certain initial values of the {\it static} parameters it is
possible that the unstable Heisenberg FP is reached (see Fig. 3 in
paper I the flow number 1, which lies on the surface separating the
attraction region of the biconical FP from the flow away solutions.
In such a case the effective exponents reach faster the asymptotic
value of the dynamical exponent $z=2.44$ since the dynamical transient
exponent $\omega_{w_\perp}=B$ is lager by  a factor of $33$ (for the
values of B see caption to Tab. \ref{tab3}).

\section{Conclusion and outlook} \label{out}

The effect of coupling the conserved magnetization parallel to the
external magnetic field to the two OPs (the components of the
alternating magnetization parallel and perpendicular to the external
magnetic field) leads to strong scaling dynamical behavior. This
means that the timescales of all dynamical quantities scale with the
same dynamical critical exponent $z=2\phi/\nu-2$. However this
scaling behavior might be hidden by nonasymtotic effects dominating
the physical accessible region when approaching the multicritical
point due to  a very small dynamical transient. This
applies in the physical interesting case $n_\|=1$ and $n_\perp=2$
where one of the timescales is almost zero and the other one almost
infinite. In consequence the magnetic transport coefficients might
show different effective behavior with temperature when
approaching the multicritical point. The dynamical amplitude
ratios might be far from their asymptotic values and show
nonuniversal behavior.

For a complete description in the whole $n_\|$-$n_\perp$-space the
model presented here has to be extended in two ways. First for the physical
case $n_\|=1$ and $n_\perp=2$ one has to introduce reversible terms
in the equations of motions. Second  one has to allow for an asymmetric
coupling to a energy like CD in addition to the magnetization.

Acknowledgement:   This
work was supported by the Fonds zur F\"orderung der
wissenschaftlichen Forschung under Project No. P19583-N20.


\begin{thebibliography}{99}
\bibitem{partI} R. Folk, Y. Holovatch, and G. Moser,  Phys. Rev. E {\bf 78}, 041125 (2008); 
arXiv:0808.0314; henceforth called paper I
\bibitem{partII} R. Folk, Y. Holovatch, and G. Moser,  Phys. Rev.E {\bf 78}, 041124 (2008); 
arXiv:0809.3146; henceforth called paper II
\bibitem{dohmjanssen77} V. Dohm and H.-K. Janssen, Phys. Rev. Lett. {\bf 39}, 946 (1977); J. Appl. Phys. {\bf 49}, 1347
(1978)
\bibitem{rem1} Since at the multicritical point two different types of fluctuation are present corresponding to the two OPs
there are also two types of slow conserved densities: (i) magnetic like and (ii) energy like. They correspond to the two
fields: (i) the magnetic field and (ii) the temperature.
\bibitem{dohmKFA} V. Dohm, {\em Report of the Kernforschungsanlage J\"ulich Nr.
1578} (1979)
\bibitem{dohmmulti83} V. Dohm in {\it Multicritical Phenomena}, ed. Plenum,
New York and London 1983 page 81
\bibitem{hahoma74} B. I. Halperin, P.C.Hohenberg, and Shang-keng Ma, Phys. Rev. B
{\bf 10}, 139 (1974); E. Brezin and C. De Dominicis, Phys. Rev. B {\bf 12}, 4954 (1975)
\bibitem{fomoprl03} R. Folk and G. Moser,  Phys. Rev. Lett. {\bf 91}, 030601 (2003)
\bibitem{phi4} D. R. Nelson, J. M. Kosterlitz, and M. E. Fisher, Phys. Rev. Lett. {\bf 33}, 813 (1974);
Y. Imry, J. Phys. C: Solid State Phys. {\bf 8}, 567 (1975); I. F. Lyuksyutov, V. L. Pokrovskii, and D. E.
Khmelnitskii, Sov. Phys. JETP {\bf 42}, 923 (1975); V. V.
Prudnikov, P. V. Prudnikov, and A. A. Fedorenko, JETP Lett. {\bf
68}, 950 (1998); P. Calabrese, A. Pelissetto, and E. Vicari, Phys.
Rev. {\bf 67}, 054505 (2003) 
\bibitem{minsub} V. Dohm, Z. Phys. B {\bf 60}, 61 (1985);
R. Schloms and V. Dohm, Europhys. Lett.,  {\bf 3}, 413 (1987); R.
Schloms and V. Dohm,  Nucl. Phys. B {\bf 328}, 639 (1989).
\bibitem{review} R. Folk and G. Moser, J. Phys. A: Math. Gen. {\bf 39}, R207 (2006)  
\bibitem{fomo04} R. Folk and G. Moser, Phys. Rev. E {\bf 69}, 036101 (2004).
\bibitem{changhoughton80} M-C. Chang and A. Houghton, Phys. Rev. B {\bf 21},
1881 (1980)
\bibitem{resum} We use the Pad\'e-Borel resummation technique, as
explained in details in paper I. For an overview see e.g. Yu.
Holovatch, V. Blavats'ka, M. Dudka, C. von Ferber, R. Folk, and T.
Yavors'kii, Int. J. Mod. Phys. B {\bf 16}, 4027 (2002).
\bibitem{Bervillier86} C. Bervillier, Phys. Rev. B {\bf 34}, 8141 (1986).
\bibitem{note1} For estimates of the marginal dimension $n_c$ see:
M. Dudka, Yu. Holovatch, and T. Yavors'kii, Acta Phys. Slovaca {\bf 52},
323 (2002).
\bibitem{Jug83} G. Jug, Phys. Rev. B {\bf 27}, 609 (1983).
\bibitem{note2} Two cautions are to be made at this point. First,
due to the special structure of the dynamical vertex functions, the
dynamical $\beta$-functions  contain only even powers of couplings
$\gamma_\perp$ and $\gamma_\|$. Therefore, to define a sign of the
coupling, one should use an additional condition, Eq. (\ref{vgpegpa}). Second, as
far as the formulas for the dynamical $\beta$-functions are
non-polynomial, one can not use familiar resummation techniques for
further numerical estimates. However one improves their convergence
in an indirect way, by using there the resummed values for the
static couplings.
\bibitem{rem2} Note that in Eq. (\ref{hbicritc}) only one CD has been taken into account. In the most general case and
especially when the specific heat is diverging one should allow for a second CD and four asymmetric couplings.
\end{thebibliography}
\end{document}